\documentclass[useAMS,usenatbib,babel]{mn2e}

\usepackage[english,english]{babel}
\usepackage{amsmath}
\usepackage{amssymb,amsfonts,textcomp}
\usepackage{array}
\usepackage{supertabular}
\usepackage{hhline}
\usepackage{hyperref}
\usepackage[usenames]{color}
\hypersetup{dvips, colorlinks=true, linkcolor=black, citecolor=black, filecolor=blue, urlcolor=blue}
\usepackage[dvips]{graphicx}

\def\gtrsim{\lower.5ex\hbox{$\; \buildrel > \over \sim \;$}}
\usepackage{graphicx}

\definecolor{grey}{rgb}{0.75,0.75,0.75}
\definecolor{Orange}{rgb}{1.0,0.5,0.15}
\definecolor{brown}{rgb}{0.7,0.25,0.0}
\definecolor{pink}{rgb}{1.0,0.5,0.5}
\definecolor{darkerred}{rgb}{0.8,0,0}
\definecolor{darkerblue}{rgb}{0,0,0.8}
\definecolor{Blue}{rgb}{0,0.08,0.65}
\definecolor{Red}{rgb}{0.65,0.08,0.05}
\definecolor{Green}{rgb}{0.15,0.45,0.25}

\begin{document}

\author[Y. Dubois et al. ]{
\parbox[t]{\textwidth}{
Yohan Dubois$^{1,2}$\thanks{E-mail: dubois@iap.fr}, Marta Volonteri$^{1,3}$, Joseph Silk$^{1,2,4}$, Julien Devriendt$^{2,5}$ \\
and Adrianne Slyz$^{2}$}
\vspace*{6pt} \\
$^{1}$ Institut d'Astrophysique de Paris, UMR 7095, CNRS, UPMC Univ. Paris VI, 98 bis boulevard Arago, 75014 Paris, France\\
$^{2}$ Sub-department of Astrophysics, University of Oxford, Keble Road, Oxford OX1 3RH\\
$^{3}$ Astronomy Department, University of Michigan, Ann Arbor, MI 48109, USA\\
$^{4}$ Department of Physics and Astronomy, The Johns Hopkins University Homewood Campus, Baltimore, MD 21218, USA\\
$^{5}$ Observatoire de Lyon, UMR 5574, 9 avenue Charles Andr\'e, Saint Genis Laval 69561, France\\
}
\date{Accepted 2014 March 2.  Received 2014 February 27; in original form 2013 December 19}

\title[BH spin in a turbulent ISM]
{Black hole evolution: II. Spinning black holes in a supernova-driven turbulent interstellar medium}

\maketitle

\begin{abstract}
{Supermassive black holes (BH) accrete gas from their surroundings and coalesce with companions during galaxy mergers, and both processes change the BH mass and spin. By means of high-resolution hydrodynamical simulations of galaxies, either idealised or embedded within the cosmic web, we explore the effects of interstellar gas dynamics and external perturbations on BH spin evolution. All these physical quantities were evolved on-the-fly in a self-consistent manner.  We use a `maximal' model to describe the turbulence induced by stellar feedback to highlight its impact on the angular momentum of the gas accreted by the BH. Periods of intense star formation are followed by phases where stellar feedback drives large-scale outflows and hot bubbles. We find that BH accretion is synchronised with star formation, as only when gas is cold and dense do both processes take place. During such periods, gas motion is dominated by consistent rotation. On the other hand, when stellar feedback becomes substantial, turbulent motion randomises gas angular momentum. However BH accretion is strongly suppressed in that case, as cold and dense gas is lacking. In our cosmological simulation, at very early times ($z>6$),
the galactic disc has not yet settled and no preferred direction exists for the angular momentum of the accreted gas, so the BH spin remains low. 
As the gas settles into a disc ($6 > z> 3$), the BH spin then rapidly reaches its maximal value. At lower redshifts ($z<3$), even when galaxy mergers flip the direction of the angular momentum of the accreted gas, causing it to counter-rotate, the BH spin magnitude only decreases modestly and temporarily. Should this be a typical evolution scenario for BH, it potentially has dramatic consequences regarding their origin and assembly, as accretion on maximally spinning BH embedded in thin Shakura-Sunyaev disc is significantly reduced.}
\end{abstract}


\section{Introduction}

Supermassive black holes (BH) are compact objects commonly observed in the centre of galaxies.
They are suspected to grow along with their host galaxies as observations suggest a strong scaling between BH masses and galaxy properties~\citep{magorrianetal98, tremaineetal02, haring&rix04}.
Gas accretion, as opposed to mergers, drives most of the mass growth of black holes at high-redshift~\citep{yutremaine2002, marconietal04, malbonetal07, shankaretal09, volonteri&ciotti13, duboisetal13, kulieretal13} either through direct accretion of cosmic filamentary gas or of star-forming clumps~\citep{bournaudetal11, dimatteoetal12, duboisetal12angmom, duboisetal13, bellovaryetal13, fengetal13}. 
This mechanism converts a fraction of the accreted rest-mass energy into effective feedback for the host galaxy, which in turn can explain the observed scaling relation between black holes and galaxies~\citep{silk&rees98, king03, wyithe&loeb03}.
Such a behaviour has been successfully implemented in modern cosmological simulations, and the impact of active galactic nucleus (AGN) feedback on galaxy properties can be dramatic~\citep{dimatteoetal05, crotonetal06, boweretal06, sijackietal07, booth&schaye09, duboisetal10, teyssieretal11, duboisetal12agnmodel}.

In thin accretion discs \citep[e.g.,][]{shakura&sunyaev73}, the radiative efficiency, i.e., the fraction of energy released and radiated during accretion events on to central BHs can be related directly to the spin magnitude, via the potential energy, $E_{\rm isco}$, of a particle on the innermost stable orbit:  $\epsilon_{\rm r}=1- E_{\rm isco}=1-\sqrt{1-2/(3r_{\rm isco})}$. Since the last stable orbit, $r_{\rm isco}$,  is closer in for BHs with high spin, corotating with the accretion disc, they liberate more energy than non-spinning or counter-rotating BHs at fixed accretion rate.
Therefore, the magnitude and relative orientation of BH spins has a direct consequence on the mass growth, as the maximum rate at which BHs accrete, the Eddington limit, is inversely proportional to radiative efficiency of the accretion process: highly spinning BHs grow more slowly than BHs with lower spin values. Consequently, the rate at which the high-redshift BHs are able to gain their mass as observed above $z>6$~\citep{fanetal06, jiangetal09, mortlocketal11, johnson2013} underscores the need to jointly study BH mass and spin evolution.

The current measurements of spins through X-ray spectroscopy still  have large uncertainties or controversial measurements~\citep[see the recent review by][and references therein]{reynoldsetal13} due to  degeneracies in the underlying models that try to reproduce the observed spectral energy distribution of AGN.
Therefore, theoretical models and numerical simulations could provide powerful tools for predicting spin values and interpreting the data more precisely.
Theoretical models~\citep{moderski1996a,moderski1996b,moderski1998} forecast that  BH spin grows to large values if the angular momentum (AM) of the accreted material onto the BH is aligned with that of the BH or has some level of coherency. A completely random distribution of the accreted gas AM is needed to drive BH spins towards zero~\citep{kingetal08, dottietal13}. In this paper, we take advantage of isolated galaxy simulations and a direct cosmological simulation performed at high spatial and mass resolution to study the effect of turbulence in the interstellar medium (ISM) on the distribution 
of gas AM. More  specifically, we investigate whether the motion of interstellar gas reaches a chaotic state such that the gas AM reorientates on a short time-scale (shorter than a Salpeter time-scale).

Some recent numerical work have been undertaken by different authors that led to divergent results:~\cite{maioetal13} use {\sc gadget} \citep{springel05} to run simulations at very high spatial resolution, and follow star formation and supernova (SN) feedback (but not BH feedback) self-consistently down to sub-pc scale, but adopt an idealised set-up, i.e. a massive circumnuclear disc within an external bulge potential. They find that, independently of resolution and feedback prescriptions, the gas that feeds the BH has a net AM aligned with that of the circumnuclear disc, despite the high level of turbulence induced by motions of gas clumps and stellar feedback. Conversely,~\cite{hopkinsetal12} use a series of nested simulations run with {\sc gadget} to track the AM of gas flowing from kpc to pc scales. They start from representative examples of gas-rich mergers and isolated, moderately bar-unstable disc simulations, and through a series of resimulations, reach a resolution limit of 0.1 pc. While they do not explicitly model stellar or BH feedback but use an effective equation of state instead, the advantage of their strategy is that the large-scale torques due to galaxy mergers and  bar instabilities are taken into account. In these simulations ,they often find misalignments between the inner $1\, \rm pc$ gas AM and that at large scales (even though they do not follow the spin evolution of a central BH) resulting from a galaxy merger.

This paper is part of a series of three papers investigating the connection between SN feedback and BH growth~(Dubois et al., in prep., paper I), SN-driven gas turbulence and BH spin evolution, (this paper, paper II), and the impact of galaxy evolution on the spin of a cosmologically representative sample of central BH ~\citep[][paper III]{duboisetal13spinlss}.
The main purpose here is to investigate the spin (and mass) evolution of a central BH in high-resolution hydrodynamical simulations, including a cosmological run. 
This allows one to assess, albeit in a specific case, the influence of galaxy mergers and simultaneously probe how gas on galaxy scales connects to the inner region surrounding the BH. The simulations have enough resolution, 10 pc, to start resolving the substructures of the ISM over several rotation periods of the galactic disc ($\sim 50-100 \, \rm Myr$ in our simulations). Moreover, we use the recent non-thermal SN feedback prescription from~\cite{teyssieretal13} which helps to expel the dense cold gas from the galaxy and maximise the level of gas turbulence present in the galaxy. The idea behind such a numerical set-up is to
provide as low a level of coherence in the AM flowing to the BH as possible. 

In Section~\ref{section:numerics} we introduce the initial conditions and the numerical models for the physics of galaxy formation, BH mass growth and its associated AGN feedback.
In Section~\ref{section:spinmodel} we provide a detailed description as to how BH spin evolves in our simulations, which depends on gas accretion and BH-BH mergers.
In Section~\ref{section:result} we detail our results for the evolution of BH spin in an isolated galaxy and for a BH with similar mass embedded within its cosmological environment.
Finally, in Section~\ref{section:conclusion} we discuss our results.

\section{Numerical set-up}
\label{section:numerics}

The simulations are run with the Adaptive Mesh Refinement code {\sc ramses} \citep{teyssier02}.
The evolution of the gas is followed using a second-order unsplit Godunov scheme for the Euler equations.
The HLLC Riemann solver with a first-order MinMod Total Variation Diminishing scheme to reconstruct the interpolated variables from their cell-centered values is used to compute fluxes at cell interfaces.
Collisionless particles (DM, star and BH particles) are evolved using a particle-mesh solver with a Cloud-In-Cell interpolation.

\subsection{Isolated galaxy}

For the isolated galaxy run, the gas is initially set in hydrostatic equilibrium within a live NFW ~\citep{navarroetal97} dark matter (DM) halo with a concentration parameter $c=3.5$. 
The total mass of gas plus DM is $M_{\rm vir}=10^{12}\, \rm M_\odot$. 
The gas fraction is 15 per cent, and the gas has some initial AM with a total spin parameter of $0.04$, slightly lower but consistent with the average spin parameter of 
cosmological dark matter halos~\citep{bullocketal01}.
When the simulation starts, the gas loses its internal pressure support due to gas cooling and a centrifugally supported disc settles in the centre of the gravitational potential well.

The mass of a DM particle is $M_{\rm DM, res}=1.7 \times 10^6 \, \rm M_\odot$, and the minimum grid cell size is of $\Delta x=9\, \rm pc$.
A central BH is inserted after 60 Myr, when the gas density has become sufficiently high to start forming stars, with an initial seed mass of $10^4 \, \rm M_\odot$. A resolution study is presented in Appendix~\ref{appendix:resolution}.

\subsection{Cosmological galaxy}

We assume a $\Lambda$CDM cosmology with total matter density $\Omega_{m}=0.3$, baryon density $\Omega_b=0.045$, dark energy density $\Omega_{\Lambda}=0.7$, amplitude of the matter power spectrum $\sigma_8=0.8285$, $n_{\rm s}=0.9635$ spectral index and  Hubble constant $H_0=68.14\, \rm km\, s^{-1} \, \rm Mpc^{-1}$ consistent with the Planck data~\citep{planckcosmo} for the initial conditions produced with~{\sc music}~\citep{hahn&abel11}.
The box size of our simulation is $L_{\rm box}=50\,  h^{-1}\, \rm Mpc$, with a coarse grid of $256^3$ DM particles corresponding to a DM mass resolution of $M_{\rm res,coarse}=8\times 10^8 \, \rm M_\odot$.
A high-resolution region is defined around a halo of $M_{\rm vir}=10^{12}\, \rm M_\odot$ at $z=2$ that contains only high-resolution DM particles within 2 $r_{\rm vir}$ ($r_{\rm vir}=100$~kpc) with mass $M_{\rm res,high}=2\times 10^5 \, \rm M_\odot$.

The mesh is refined up to $\Delta x=8.7$~pc (maximum level of refinement equals 21 at redshift $z=3$) using a quasi-Lagrangian strategy: we refine when the mass in a cell  
becomes larger than 8 times the mass of our high resolution DM particles.
The minimum cell size is kept roughly constant in physical size with redshift, i.e. an additional level of refinement is added every $a_{\rm exp}=n\times0.1$ (where $n=1,2,4,8$ and $a_{\rm exp}$ is the expansion factor of the universe) up to level 21 at $a_{\rm exp}=0.33$. 
The BH seed mass used in the cosmological simulation is larger than for the isolated runs, at $10^5 \, \rm M_\odot$.

Such a high spatial resolution for a cosmological zoomed halo evolved over several billion years places this simulation amongst the most resolved state-of-the hydrodynamical cosmological simulations, including complex sub-grid processes, to date~\cite[e.g.][for comparison]{bellovaryetal13, ceverinoetal13, roskaretal14, hopkinsetal14}.
We call this simulation and its various flavours (resolution, physics) the {\sc Seth} simulation suite\footnote{The name is a reference to the Egyptian god of desert, storms, disorder and violence.
It is also a reference to the code employed for the simulations, {\sc ramses}, and suite of simulations with similar (as well as much higher) resolution performed for a lower mass galaxy at high redshift, the {\sc Nut} suite~\citep{powelletal11}. }.

\subsection{Physics of galaxy formation}

Gas is allowed to cool by H and He cooling with a contribution from metals using a~\cite{sutherland&dopita93} model for temperatures above $T_0=10^3$~K, which is the minimum temperature  we allow the gas to reach through radiative losses.
Heating from a uniform UV background takes place after redshift $z_{\rm reion}=10$ following~\cite{haardt&madau96}.
Metallicity is modelled as a passive variable for the gas advected with the flow (whose composition is assumed to be solar) and is altered by gas ejecta from SN explosions and stellar winds.
We assume a zero initial metallicity.
The gas follows the equation of state (EoS) of an ideal monoatomic gas $P=(\gamma -1)e$, to relate the gas pressure $P$ to the internal energy density $e$, with $\gamma=5/3$.

The star formation process is modelled with a Schmidt law:
$\dot \rho_*= \epsilon_* {\rho / t_{\rm ff}}\, ,$ where $\dot \rho_*$ is the star formation rate density, $\epsilon_*$ the constant star formation efficiency, and $t_{\rm ff}$ the local free-fall time of the gas.
We choose a low star formation efficiency $\epsilon_*=0.02$ consistent with observations of giant molecular clouds~\citep{krumholz&tan07} and surface density relations of galaxies~\citep{kennicutt98}.
Star formation is allowed in regions exceeding a gas density threshold of $n_0=250\, \rm H\, cm^{-3}$ and for gas with a temperature given by the polytropic EoS described below.
The minimum mass of a star particle is $m_{\rm s, res}=n_0\Delta x^3\simeq4\times 10^3 \, \rm M_\odot$ at our spatial resolution $\Delta x=9$ pc.
The gas pressure is artificially enhanced above $\rho > \rho_0$ assuming a polytropic EoS $T=T_0(\rho/\rho_0)^{\kappa-1}$ with polytropic index $\kappa=2$ to avoid 
spurious gas fragmentation: we always resolve the Jeans length with at least four cells.
Feedback from massive stars is taken into account assuming a Salpeter initial mass function with $\eta_{\rm SN}=0.1$ of the mass fraction of stars ending up as type II SN and releasing $10^{50} \, \rm erg\, M_\odot^{-1}$. We use the feedback model introduced in \cite[][similar in spirit to the delayed cooling prescription from~\citealp{stinsonetal06}]{teyssieretal13}, where energy is released both as a thermal component and a tracer,``non-thermal'' component, which is passively advected with the flow. This passive non-thermal energy decays on a typical time-scale of $t_{\rm diss}=0.8$~Myr, and gas cooling is prevented until the velocity dispersion associated with this non-thermal energy component becomes larger than $\sigma_{\rm NT}=50 \, \rm km.s^{-1}$. The values of $t_{\rm diss}$ and $\sigma_{\rm NT}$ are obtained by imposing that $t_{\rm diss}=\lambda_{\rm J}/\sigma_{\rm NT}$, where $\lambda_{\rm J}=4\times \Delta x$ is the resolved Jeans length. In isolated dwarf galaxies, this numerical implementation of SN feedback allows for the destruction of cold dense star-forming clouds, and for an efficient driving of large-scale galactic winds and turbulence in the ISM as shown in~\cite{teyssieretal13}.

An alternative modelling of SN feedback (which drives large-scale winds but does not manage to destroy completely the cold clumps of gas) through kinetic Sedov blast wave explosions~\citep{dubois&teyssier08winds} and its impact on the BH spin evolution is presented in the Appendix~\ref{appendix:snkinetic}. Its relative impact on the BH growth compared to the reference case is the topic of Paper I.

\subsection{Model for BH growth and AGN feedback}

We use the same ``canonical'' BH model employed in~\cite{duboisetal12agnmodel}.
BH are created at loci where gas density is larger than the density threshold for star formation $\rho_0$ with an initial seed mass of $10^4$ or $10^5\, \rm M_\odot$ (respectively for the isolated and the cosmological simulations).
In order to avoid the formation of multiple BHs in the same galaxy, they are not allowed to form at distances smaller than 50 kpc (comoving) from any other BH particle.
The accretion rate onto BH follows the Bondi-Hoyle-Lyttleton~\citep{bondi52} rate
$\dot M_{\rm BH}=4\pi \alpha G^2 M_{\rm BH}^2 \bar \rho / (\bar c_s^2+\bar u^2) ^{3/2},$
where $M_{\rm BH}$ is the BH mass, $\bar \rho$ is the average gas density, $\bar c_s$ is the average sound speed, $\bar u$ is the average gas velocity relative to the BH velocity, and $\alpha$ is a dimensionless boost factor with $\alpha=(\rho/\rho_0)^2$ when $\rho>\rho_0$ and $\alpha=1$ otherwise~\citep{booth&schaye09} in order to account for our inability to capture the colder and higher density regions of the ISM which are suppressed by the presence of the polytropic EoS.
The effective accretion rate onto BH is capped at the Eddington accretion rate:
$\dot M_{\rm Edd}=4\pi G M_{\rm BH}m_{\rm p} / (\epsilon_{\rm r} \sigma_{\rm T} c),$
where $\sigma_{\rm T}$ is the Thompson cross-section, $c$ is the speed of light, $m_{\rm p}$ is the proton mass, and $\epsilon_{\rm r}$ is the radiative efficiency, assumed to be proportional to the spin of the BH with $\epsilon_{\rm r}=1- E_{\rm isco}=1-\sqrt{1-2/(3r_{\rm isco})}$.

In order to avoid spurious oscillations of the BH in the gravitational potential well due to external perturbations and finite resolution effects, we introduce a drag force that mimics the dynamical friction exerted by the gas onto a massive particle.
This dynamical friction is proportional to $F_{\rm DF}=f_{\rm gas} 4 \pi \alpha \rho (G M_{\rm BH}/\bar c_s)^2$, where $f_{\rm gas}$ is a fudge factor whose value is between 0 and 2 and is a function of the mach number ${\mathcal M}=\bar u/\bar c_s<1$~\citep{ostriker99, chaponetal13}, and where we introduce the boost factor $\alpha$ for reasons  
previously stated.

Two BH are allowed to merge when the distance separating the two particles is smaller than $4\times \Delta x$ and when their relative velocity is smaller than the escape velocity of the binary.

The AGN feedback is a combination of two different modes, the so-called \emph{radio} mode operating when $\chi=\dot M_{\rm BH}/\dot M_{\rm Edd}< 0.01$ and the \emph{quasar} mode active otherwise.
The quasar mode corresponds to an isotropic injection of thermal energy into the gas within a sphere of radius $\Delta x$, with an energy deposition rate: $\dot E_{\rm AGN}=\epsilon_{\rm f} \epsilon_{\rm r} \dot M_{\rm BH}c^2$,
where $\epsilon_{\rm f}=0.15$  is a free parameter chosen to reproduce the $M_{\rm BH}$-$M_{\rm b}$, $M_{\rm BH}$-$\sigma_{\rm b}$, and BH density in our local Universe (see \citealp{duboisetal12agnmodel}).
At low accretion rates, on the other hand, the radio mode deposits the AGN feedback energy into a bipolar outflow with a jet velocity of $10^4\,\rm km.s^{-1}$ into a cylinder with a cross-section of radius $\Delta x$ and height $2 \, \Delta x$ following~\cite{ommaetal04} (more details about the jet implementation are given in~\citealp{duboisetal10}).
The efficiency of the radio mode is larger with $\epsilon_{\rm f}=1$.

\section{Model of BH spin evolution}
\label{section:spinmodel}

\subsection{Gas accretion}

The magnitude of BH spins is allowed to change by accretion of gas through the following expression~\citep{bardeen70}:
\begin{equation}
\label{aspinup}
a^{\rm n+1}={1\over 3} {r_{\rm isco}^{1/2}\over M_{\rm ratio}}\left [ 4- \left( 3 {r_{\rm isco}\over M_{\rm ratio}^2}-2 \right )^{1/2}\right ]\, ,
\end{equation}
where $M_{\rm ratio}=M^{\rm n+1}_{\rm BH}/M^{\rm n}_{\rm BH}$, the n superscript stands for the values at time $t_{\rm n}$, and $R_{\rm isco}$ is the radius of the innermost stable circular orbit (ISCO) defined as (in reduced units):
\begin{equation}
r_{\rm isco}=R_{\rm isco}/R_{\rm g}=3+Z_{2}\pm [(3-Z_1)(3+Z1+2Z_2)]^{1/2}\, , 
\end{equation}
where the gravitational radius $R_{\rm g}$ is defined as half of the Schwarzschild radius of the BH, $R_{\rm BH}$, and $Z_1$ and $Z_2$ are:
\begin{eqnarray}
Z_1&=&1+(1-a^2)^{1/3}[(1+a)^{1/3}+(1-a)^{1/3}] \, , \\
Z_2&=&(3a^2+Z_1^2)^{1/2}\, .
\end{eqnarray}
For spins co-rotating with their accretion disc, $1\le r_{\rm isco}<6$, while for spins counter-rotating with their accretion disc, $6<r_{\rm isco}\le 9$, and $r_{\rm isco}=6$ for non-spinning BH.

Equation~\ref{aspinup}, governing the evolution of the BH spin magnitude through direct accretion of gas, assumes that BH spin and disc AM are perfectly aligned (or anti-aligned), so that the direction of the angular momentum vector does not change, but only its magnitude. In the most general case, however, misalignments occur. In this case, the accretion disc experiences a torque due to the Lense-Thirring effect that causes the accretion disc to precess about the spin axis of the BH. For large enough viscosities the innermost parts of the disc are forced to rotate within the equatorial plane of the BH and a warped disc is created. The result of the Lense-Thirring precession is that the BH and disc AM end up being aligned or anti-aligned with the total AM. We can define the total AM of the system $\{$disc+BH$\}$ as $\bmath{J}_{\rm tot}=\bmath{J}_{\rm d}+\bmath{J}_{\rm BH}$. The values of the angle $\theta$ between $\bmath{J}_{\rm BH}$ and $\bmath{J}_{\rm d}$ are between $-1\le \cos \theta\le1$ where the two extrema correspond to anti-aligned and aligned cases respectively. 
As shown by \cite{kingetal05} the alignment process can be understood as a simple vectorial sum: since $\bmath{J}_{\rm tot}=\bmath{J}_{\rm d}+\bmath{J}_{\rm BH}$, the final configuration is either aligned with $\bmath{J}_{\rm d}$ or $\bmath{J}_{\rm BH}$ depending on whether the former or the latter dominates the sum, therefore, effectively, both the direction of the BH spin and the disc AM evolve towards the equilibrium configuration. Whether the BH  mostly (anti)aligns with the disc AM or vice-versa depends therefore on the relative magnitude of the two vectors, and on the initial configuration. The case for anti-alignment of the BH with the disc requires that~\citep{kingetal05}:
\begin{equation}
\label{costheta}
\cos \theta < - {J_{\rm d}\over 2 J_{\rm BH}}\, , 
\end{equation}
thus, for $\cos \theta\ge 0$, they always align, while for $\cos \theta < 0$, they eventually anti-align if the ratio $J_{\rm d}/J_{\rm BH}$ is sufficiently small compared to $\cos \theta$.

In summary, we use Equation~\ref{costheta} to decide whether the inner parts of the accretion disc are co- ou counter-rotating with the BH, we then use Equation~\ref{aspinup} to update the spin magnitude, and the vectorial sum $\bmath{J}_{\rm tot}=\bmath{J}_{\rm d}+\bmath{J}_{\rm BH}$ to obtain the new direction of the BH AM.

One difficulty we experience is in accessing  the magnitude of $\bmath{J}_{\rm d}$ as we cannot resolve the accretion disc in our simulations.  We therefore assume the tilted solution for a  thin accretion disc~\citep{shakura&sunyaev73,Scheuer1996,NatarajanPringle1998,Perego2009}, characterized by a viscosity $\nu_1=\alpha_t c_s^2 /(GM_{\rm BH}/r^3)^{1/2}$, where $\alpha_t$ is a parameter $<1$, and $c_s$ is the sound speed.  In a misaligned disc under the effect of  Lense-Thirring precession a  natural scale is the warp radius, that marks the transition between an equatorial inner disc and a misaligned outer disc: the direction of the AM of the inflowing material changes direction as it passes through the warp. Only material within the warp radius can effectively transfer its AM to the BH~\citep{volonterietal07}. This is the relevant scale to estimate whether alignment or anti-alignment occur. 
In a~\cite{shakura&sunyaev73}  thin accretion disc, one can write the ratio of the warp to Schwarzschild radius as: 
\begin{equation}
{R_{\rm warp}\over R_{\rm BH}}\simeq 6.4 \times 10^3 a^{5/8} M_{\rm BH, 8}^{1/8} \left( \epsilon_{\rm r,01} \over \chi \right)^{1/4} \left ({\nu_2\over \nu_1}\right )^{-5/8} \alpha_{\rm t,01}^{-1/2}\, , 
\end{equation}
where $\epsilon_{\rm r,01}$ is the radiative efficiency normalised to $0.1$, $\nu_1$ and $\nu_2$ are the kinematic viscosities horizontal and perpendicular to the equatorial plane of the disc: $\nu_2$ is the viscosity responsible the warp propagation, while $\nu_1$ is the viscosity responsible for driving accretion and transferring AM. 
We choose a value for $\alpha_{\rm t,01}\equiv\alpha_{\rm t}/0.1=1$~\citep{kingetal07}, and the ratio of $\nu_2/\nu_1=2(1+7\alpha_{\rm t})/(4+\alpha_{\rm t}^2)/\alpha_{\rm t}^2$ is of the form given by~\cite{ogilvie99} (equation 145). 
With these choices, $(\nu_2/\nu_1)\sim 85$  (note that for small $\alpha_t$,  $ \nu_2\sim \nu_1/\alpha_t^2$) and: 
\begin{equation}
\label{rwarp}
{R_{\rm warp}\over R_{\rm BH}}\simeq 4 \times 10^2 a^{5/8} M_{\rm BH, 8}^{1/8} \left( \epsilon_{\rm r,01} \over \chi \right)^{1/4} \left ({\nu_2/\nu_1\over 85}\right )^{-5/8} \alpha_{\rm t,01}^{-1/2}  \,. 
\end{equation}

The BH AM is simply $J_{\rm BH} = a M_{\rm BH}^{3/2} R_{\rm BH}^{1/2}$, and the disc AM can now be expressed as $J_{\rm d} \sim M_{\rm d}(R_{\rm warp}) M_{\rm BH} ^{1/2} R_{\rm warp}^{1/2}$. The disc mass within $R_{\rm warp}$ is $M_{\rm d}(R_{\rm warp})=\dot M t_{\nu_1}(R_{\rm warp})$ where $t_{\nu_1}$ is the viscous timescale for radial propagation:
\begin{equation}
{t_{\nu_1}}=5.3\times 10^5 a^{7/8} M_{\rm BH, 8}^{11/8} \left( \epsilon_{\rm r,01} \over \chi \right)^{3/4} \left ({\nu_2\over \nu_1}\right )^{-7/8} \alpha_{\rm t}^{-3/2} {\rm yr}\, 
\end{equation}
$$\sim 3.4 \times 10^5 a^{7/8} M_{\rm BH, 8}^{11/8}  \left( \epsilon_{\rm r,01} \over \chi \right)^{3/4} \left ({\nu_2/\nu_1\over 85}\right )^{-7/8} \alpha_{\rm t,01}^{-3/2} {\rm yr}.$$

The ratio of disc to BH AM can then be written as:
\begin{equation}
\label{jdjbhratio}
{J_{\rm d}\over2 J_{\rm BH}}\simeq {M_{\rm d}(R_{\rm warp}) \over  a M_{\rm BH}} \left ({R_{\rm warp}\over R_{\rm BH}}\right )^{1/2}\,
\end{equation}
$$ \sim 6.8\times 10^{-2}  \left( \chi \over \epsilon_{\rm r,01} \right)^{1/8} M_{\rm BH, 8}^{23/16} a^{3/16} \alpha_{\rm t,01}^{-7/4}\left ({\nu_2/\nu_1\over 85}\right )^{-19/16}.$$
Equation~(\ref{jdjbhratio}) allows one to evaluate if anti-alignment occurs, and provides a value for the magnitude of $\bmath{J}_{\rm d}$.
We refer the reader to \cite{dottietal13} for a detailed discussion of how the ratio $J_{\rm d}/ J_{\rm BH}$ determines the behaviour of alignment and spin evolution. 

For the direction of $\bmath{J}_{\rm d}$, we assume that the accretion disc is aligned with the gas AM, $\bmath{J}_{\rm g}$, measured in the surroundings of the BHs as extracted directly in the simulation (on $\Delta x=10\, \rm pc$ scale).
We update the orientation of the BH AM after one accretion event of $\Delta t=t_{\nu_1}$ with  $\bmath{J}_{\rm BH}^{\rm n+1}= \bmath{J}_{\rm tot}= \bmath{J}_{\rm BH}^{\rm n} + J_{\rm d}^{\rm n} \bmath{j}_{\rm g}^{\rm n}$ for the alignment case, where the small $\bmath{j}$ stands for AM unit vectors, and where $J_{\rm d}^{\rm n}$ is provided by equation (\ref{jdjbhratio}).
If the criterion for anti-alignment is met, $\bmath{J}_{\rm BH}^{\rm n+1}$ becomes $-\bmath{J}_{\rm tot}$ if $\bmath{J}_{\rm tot}.\bmath{J}_{\rm BH}^{\rm n}>0$, this occurs when the total AM dominated by the BH  \citep[case b in Fig.~1 of][] {kingetal05}. Otherwise, if the anti-alignment condition is satisfied and $\bmath{J}_{\rm tot}.\bmath{J}_{\rm BH}^{\rm n}<0$, then the total AM is dominated by the disc and it is anti-aligned with respect to that of the BH, therefore  $\bmath{J}_{\rm BH}^{\rm n+1}=\bmath{J}_{\rm tot}$ \citep[case d in Fig.~1 of][]{kingetal05}. Since each timestep $\Delta t=t_{\nu_1}$  is defined as the time needed for viscosity to transport in the radial direction the material within the warp, this corresponds to the time needed to ``swallow" this part of the accretion disc.

For sufficiently large accretion rates, accretion discs can become unstable against their own gravity and  fragment into gas clumps~\citep[e.g.][]{kolykhalov&sunyaev80, pringle81,goodman&tan04}. The criterion for stability given by the Toomre parameter $Q \sim c_{\rm s}\Omega/(\pi G\Sigma)$, where $\Omega$ is the Keplerian velocity and $\Sigma$ the gas surface density, provides the radius $R_{\rm sg}$ below which the accretion disc is stable. For the standard $\alpha$-disc external region, where pressure is dominated by the gas pressure and opacity is given by free-free transitions \citep{shakura&sunyaev73}:
\begin{equation}
\label{rsg}
{R_{\rm sg}\over R_{\rm BH}}\simeq 5 \times 10^2 \alpha_{\rm t, 01}^{28/45} M_{\rm BH, 8}^{-52/45} \left( \epsilon_{\rm r,01} \over \chi \right)^{22/45} \, , 
\end{equation}
and the mass stable against fragmentation is~\citep[see][]{dottietal13}
\begin{equation}
\label{msg}
M_{\rm sg} \simeq 6 \times 10^5 \alpha_{\rm t,01}^{-1/45} M_{\rm BH, 8}^{34/45} \left( \chi \over \epsilon_{\rm r,01} \right)^{4/45} \, \rm M_\odot\, .
\end{equation}
We include the effects of self-gravity when $R_{\rm sg}< R_{\rm warp}$. In this case, the outer region of the disc is subject to fragmentation, while the central region as a whole aligns (or anti-aligns) in the equatorial plane of the BH.  Then, the AM accreted onto the BH is the form of equation~(\ref{jdjbhratio}) with $R_{\rm warp}$ and $M_{\rm d}(R_{\rm warp})$ replaced by $R_{\rm sg}$ and $M_{\rm sg}$ respectively. Thus, we repeat the same steps as for the non self-gravitating case for the alignment (anti-alignment) criterion with the AM calculated with $R_{\rm sg} $ and $M_{\rm sg} $.
In contrast to \cite{kingetal08}, who assume that  self-gravity causes chaos in the AM distribution, we argue that the accreted gas conserves the AM direction it possessed when entering the accretion flow, which is the direction measured from the simulation by definition.
As a final note, the effects of self-gravity become important when  $R_{\rm sg}< R_{\rm warp}$, and through Eq.~(\ref{rwarp}) and~(\ref{rsg}), the critical BH mass above which the BH cannot stabilise anymore the disc against fragmentation is:
\begin{equation}
\label{rwrsg2}
M_{\rm BH, 8} > 1.15 \, a^{- {225\over 461}} \left ( {\epsilon_{\rm r,01} \over \chi} \right )^{86 \over 461} \left ({\nu_2/\nu_1\over 85}\right )^{225 \over 461}   \alpha_{\rm t, 01}^{404 \over 461}\, .
\end{equation}
Similar expressions hold when using the middle region solutions for the $\alpha$-disc \citep[where pressure is dominated by the gas pressure and opacity by electron scattering][]{shakura&sunyaev73}. In this case ${R_{\rm sg}/ R_{\rm BH}}\simeq 10^3 \alpha_{\rm t, 01}^{14/27} M_{\rm BH, 8}^{-26/27} \left( \epsilon_{\rm r,01} / \chi \right)^{8/27}$ and the radius where self-gravity truncates the disc becomes smaller than the warp radius when $M_{\rm BH, 8} > 2.23 \, a^{- {135\over 269}} \left ( {\epsilon_{\rm r,01} / \chi} \right )^{10 \over 269} \left [{(\nu_2/\nu_1)/ 85}\right ]^{135 \over 269}   \alpha_{\rm t, 01}^{220 \over 269}\, $.

Our implementation is based on analytical studies, as we cannot self-consistently include an accretion disc (on milli-pc scales) in a galaxy formation simulation. Regarding numerical studies of BH spin-disc alignment we recall that \cite{2005ApJ...623..347F}, \cite{2007ApJ...668..417F}, and \cite{2011ApJ...730...36D} find no alignment in their three-dimensional general relativistic magnetohydrodynamic simulations of relatively thick, low-accretion rate tilted discs, where however alignment is not expected to occur, as the typical accretion rate is sufficiently low that alignment would not occur in a Hubble time \citep{2011ApJ...730...36D}.  Recently \cite{sorathia2013} study the alignment process in magnetohydrodynamic simulations of relatively thin disks and find instead that the BH and the disc progressively align, inside-out. While, at the time being, it is not trivial to derive an analytical description of their simulations accurate enough to be explicitly included in our simulations, we use an extrapolation of their order-of-magnitude estimates to obtain, in turn, a qualitative validation of  our methodology. We consider their suggestion that the inclination transition radius (roughly analogue to what we call warp radius here) can be expressed as $R_{\rm T}\sim 0.5 \Phi a^{2/3} (H/R_{\rm T})^{-4/3} R_{\rm BH}$, where $1<\Phi<\alpha^{-2/3}$ and $(H/R_{\rm T})$ is the disc aspect ratio. We also assume that the alignment timescale is of order the alignment propagation radius at $R_{\rm T}$ \citep[Eq.~13 in][]{sorathia2013}. Assuming the  middle region solutions for the $\alpha$-disc we find an upper limit to the alignment timescale of $\sim 10^5 a^{7/8} M_8^{11/8} \alpha^{-15/4} \left( \chi \over \epsilon_{\rm r,01} \right)^{-3/4}\Phi^{45/16}$ yr by minimising the propagation rate of the alignment front. While this alignment timescale is very similar to our previous estimates and with the implementation in {\sc ramses}, since the estimate we obtain is approximate, we discuss in Appendix~\ref{appendix:spinalign} how our results would be modified if the alignment timescale is much longer than these estimates suggest.

The change of spin during BH coalescence is tracked in our simulations (see paper III for details).
However, there are no BH-BH mergers in our zoom cosmological simulation: a few BHs are identified within 10 per cent of the halo virial radius (typically the size extent of the galaxy), but none are close enough to merge as one BH.

\subsection{Updating spins}

We assume that the initial spin of BHs is zero at creation.
We further assume that the thin disc solution for spin evolution applies at any accretion rate over Eddington $\chi$.
It seems to be a fairly good approximation as BH spins mostly change their value by accretion during strong accretion events (of the order of $\chi \lesssim 1$).
The maximum spin value that we authorise is $a_{\rm max}=0.998$, due to the capture by the central BH of radiated photons emitted from the accretion disc~\citep{thorne74}. In the case of thick discs, however, we note that~\citep{gammieetal04} find in their numerical simulations that  $a_{\rm max}\sim0.9-0.93$, taking into account the magneto-hydrodynamical effects on the transfer of gas AM, while~\cite{sadowskietal11} suggest that super-critical accretion discs allow a maximal spin up to  $a_{\rm max}=0.9994$ in low-viscosity, $\alpha_{\rm t}=0.01$, accretion discs.

As we treat the spin evolution on-the-fly, radiative efficiencies, and, thus, Eddington accretion rate and bolometric luminosities vary with the amplitude of the BH spin.
Also, the jet of the radio mode of AGN feedback is launched along the spin axis of the BH.

\begin{figure}
  \centering{\resizebox*{!}{6cm}{\includegraphics{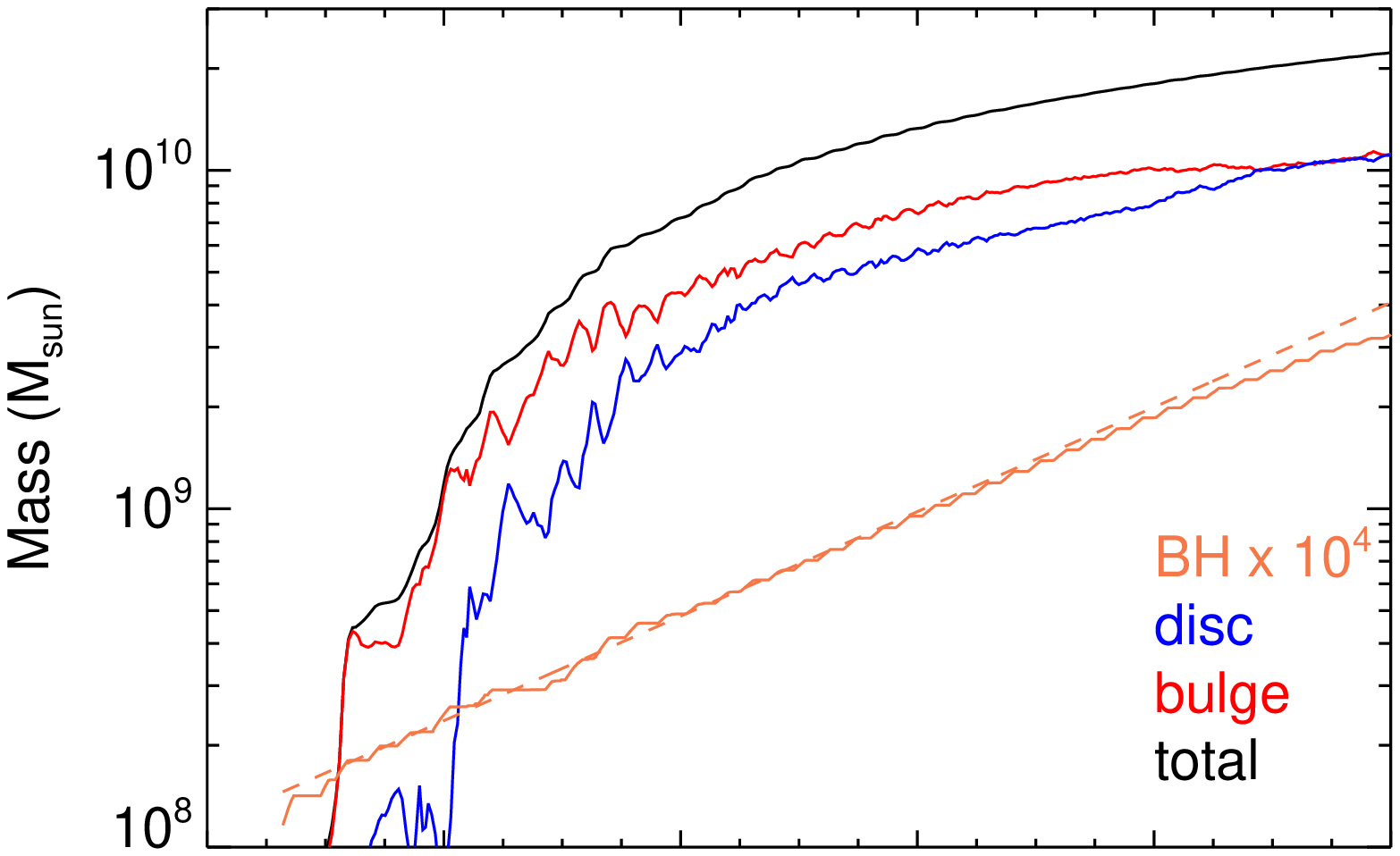}}\vspace{-1.5cm}}
  \centering{\resizebox*{!}{6cm}{\includegraphics{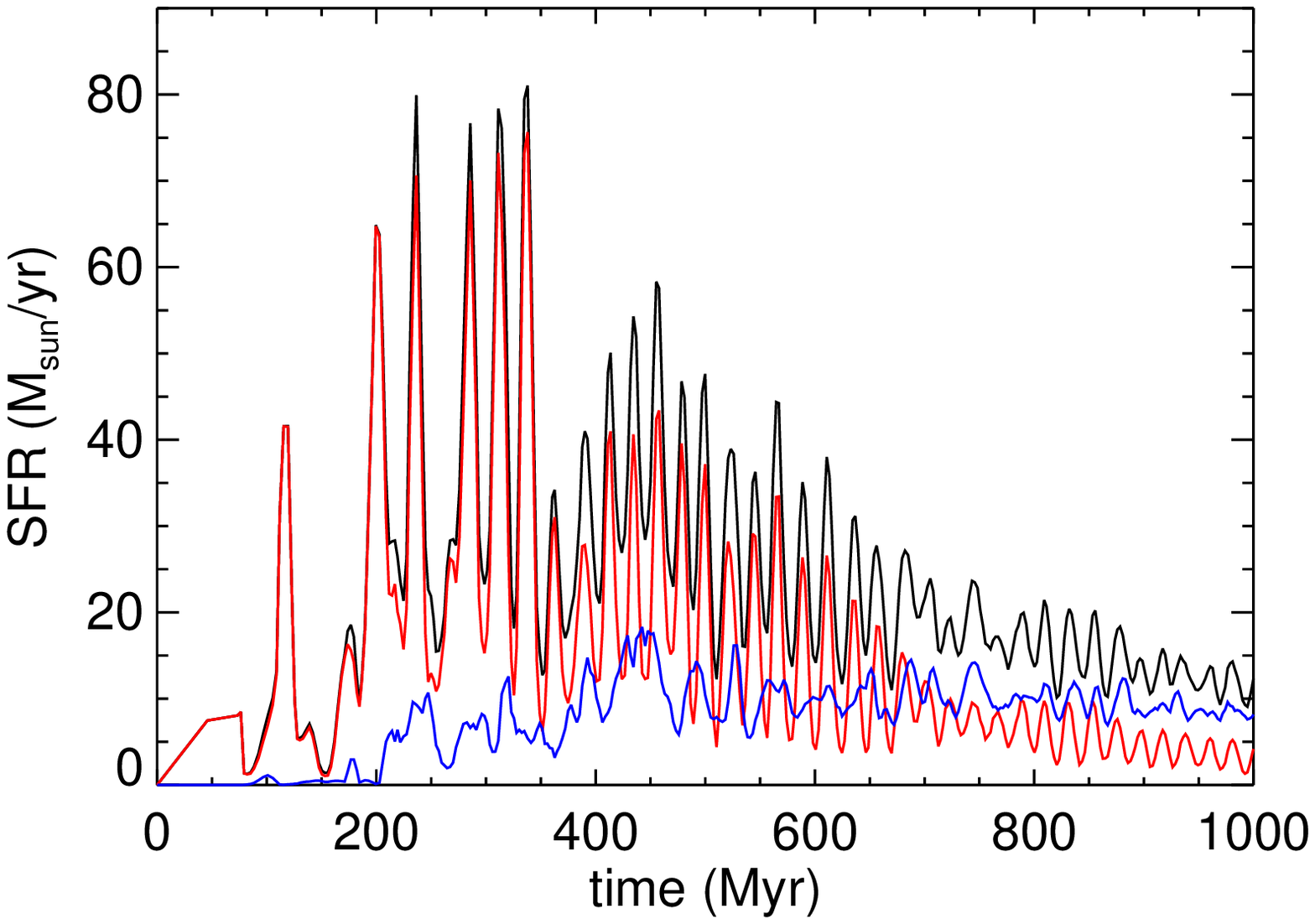}}}
  \caption{Isolated galaxy simulation. \emph{Top panel}: Total stellar mass (black), stellar mass within $r\le 500 \, \rm pc$ around the centre of the simulation box (red), disc (total minus bulge in blue), and $10^4$ times the BH mass (orange) as a function of time. The initial seed BH mass ($\times 10^4$) growing at half the Eddington rate of a maximum spinning BH ($\epsilon_{\rm r}=0.32$) is represented by the orange dashed line. \emph{Bottom panel}: Star formation rate as a function of time for the entire galaxy (black), for the stars within the bulge $r\le 500 \, \rm pc$ (red), and within the disc (blue). The SFR in the bulge shows large variations due to the regular outflows produced by SN explosions, while the SFR is more steady in the disc. The BH grows in synchronisation with the peaks of the bulge SFR.}
    \label{fig:sfrisolated}
\end{figure}

\begin{figure}
  \centering{\resizebox*{!}{4.1cm}{\includegraphics{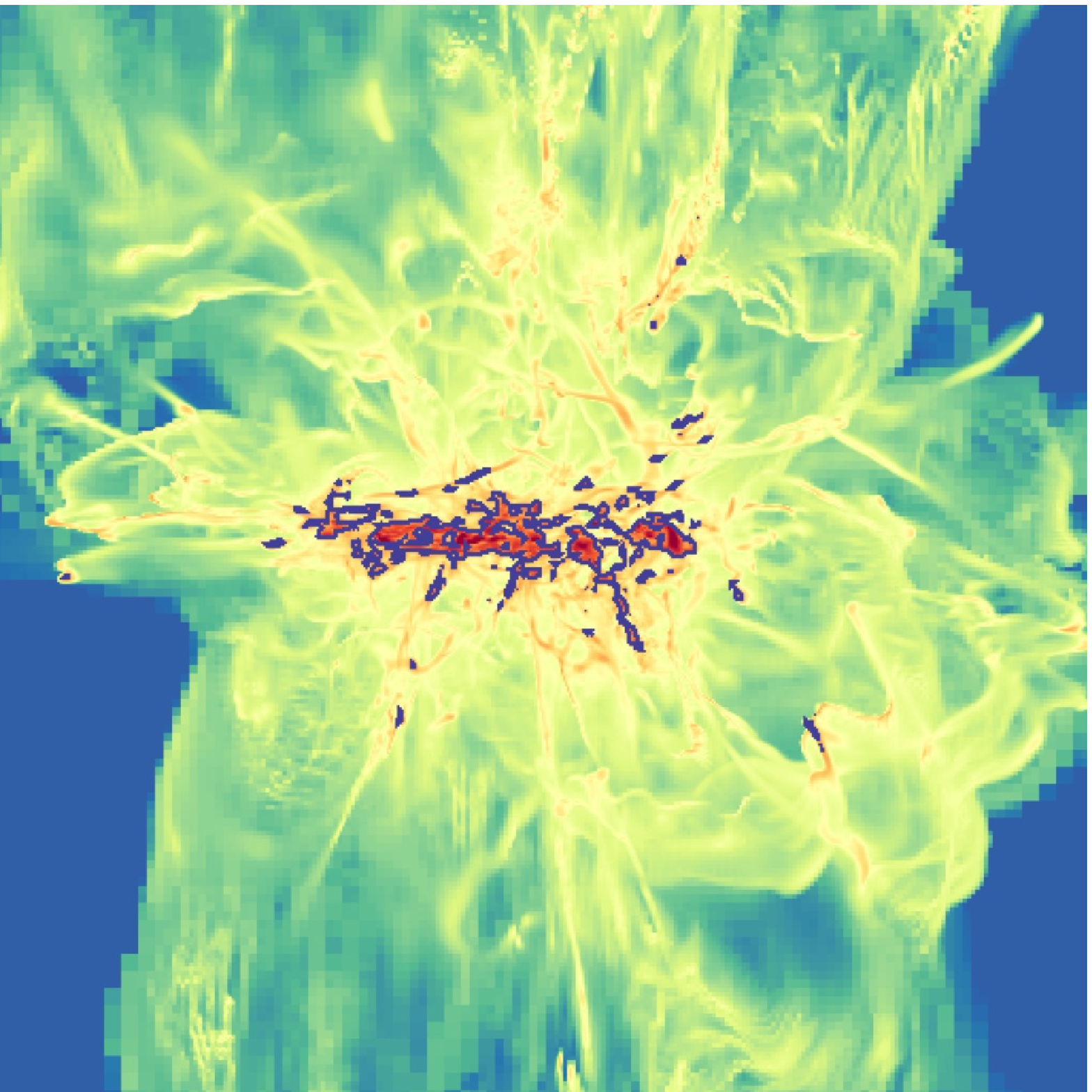}}}
  \centering{\resizebox*{!}{4.1cm}{\includegraphics{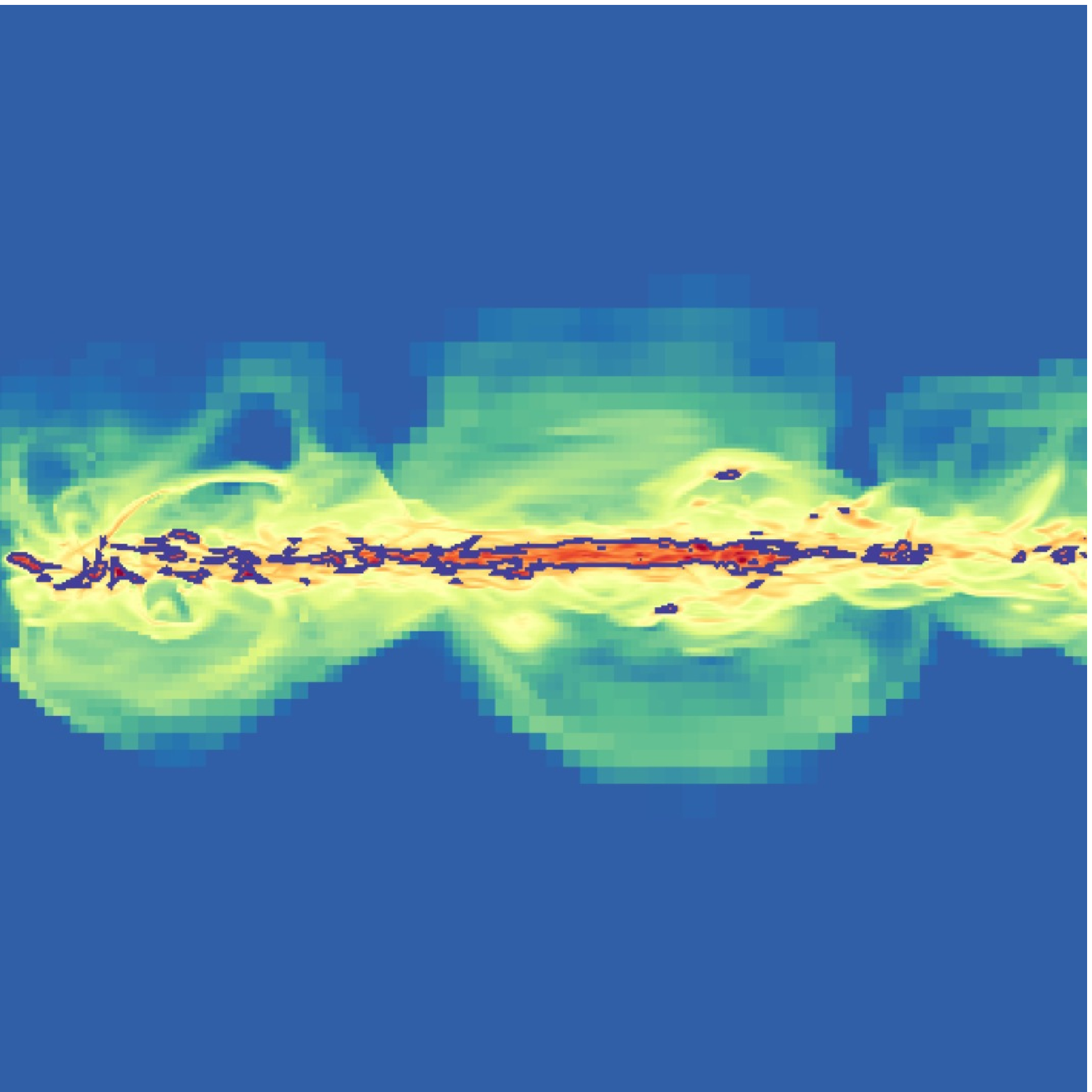}}}
  \caption{Isolated galaxy simulation.  Edge-on view of the gas density of the galaxy at $t=250$ Myr (left) and $t=1000$ Myr (right).  Image size is 5 kpc. The colour table is from $0.1 \, \rm H\, cm^{-3}$ (blue) to $10^4 \, \rm H\, cm^{-3}$ (red). The blue contours surround the regions of star formation with gas density above $250 \, \rm H\, cm^{-3}$. At early times, the strong activity of star formation produces large-scale outflows that rise up to 5 kpc above the disc and fall back into a galactic fountain. At late times, the more quiescent activity leads to the formation of hot bubbles of a few $100\, \rm pc$ size. }
    \label{fig:outflow}
\end{figure}

\begin{figure*}
  \centering{\resizebox*{!}{5.6cm}{\includegraphics{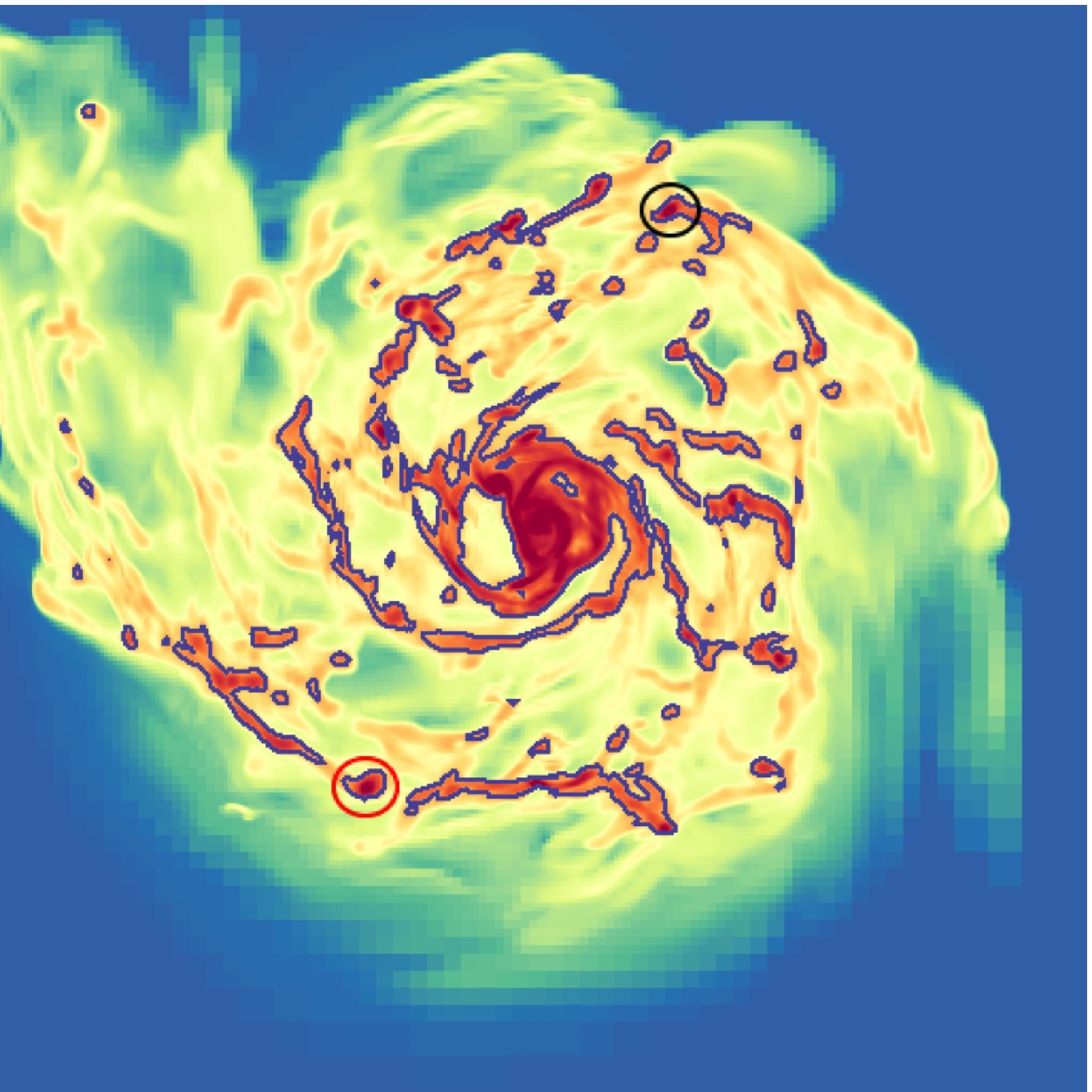}}}
  \centering{\resizebox*{!}{5.6cm}{\includegraphics{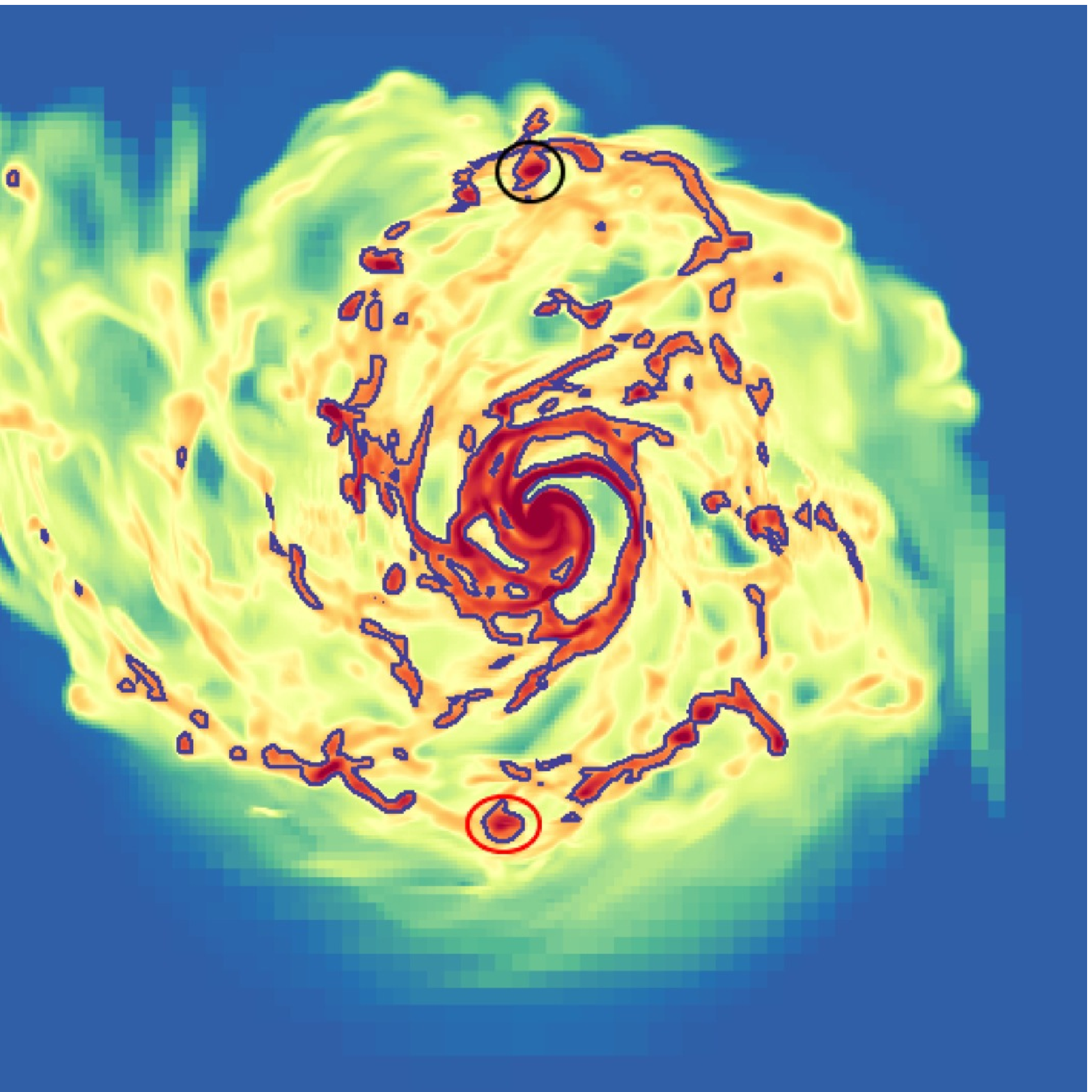}}}
  \centering{\resizebox*{!}{5.6cm}{\includegraphics{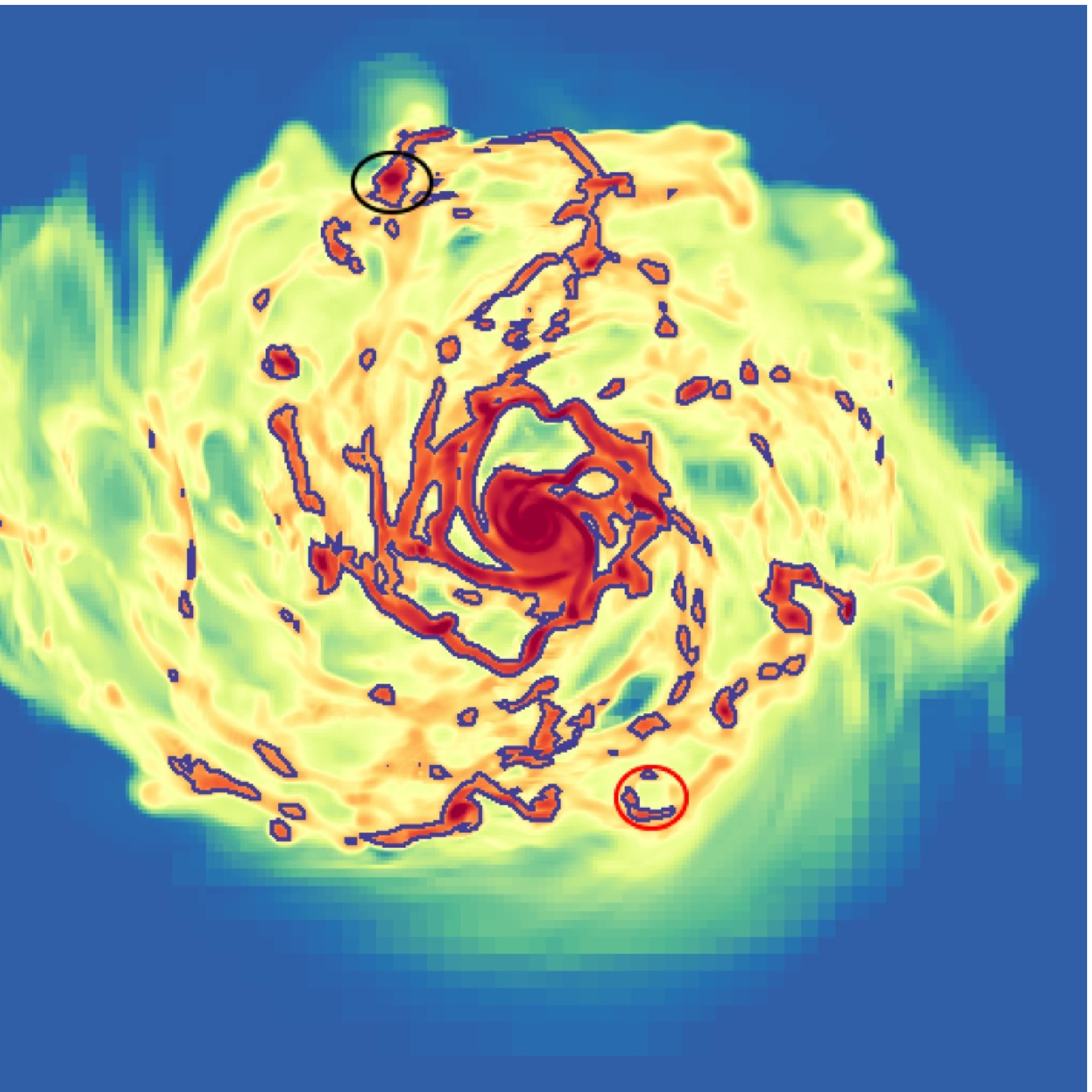}}}
  \centering{\resizebox*{!}{5.6cm}{\includegraphics{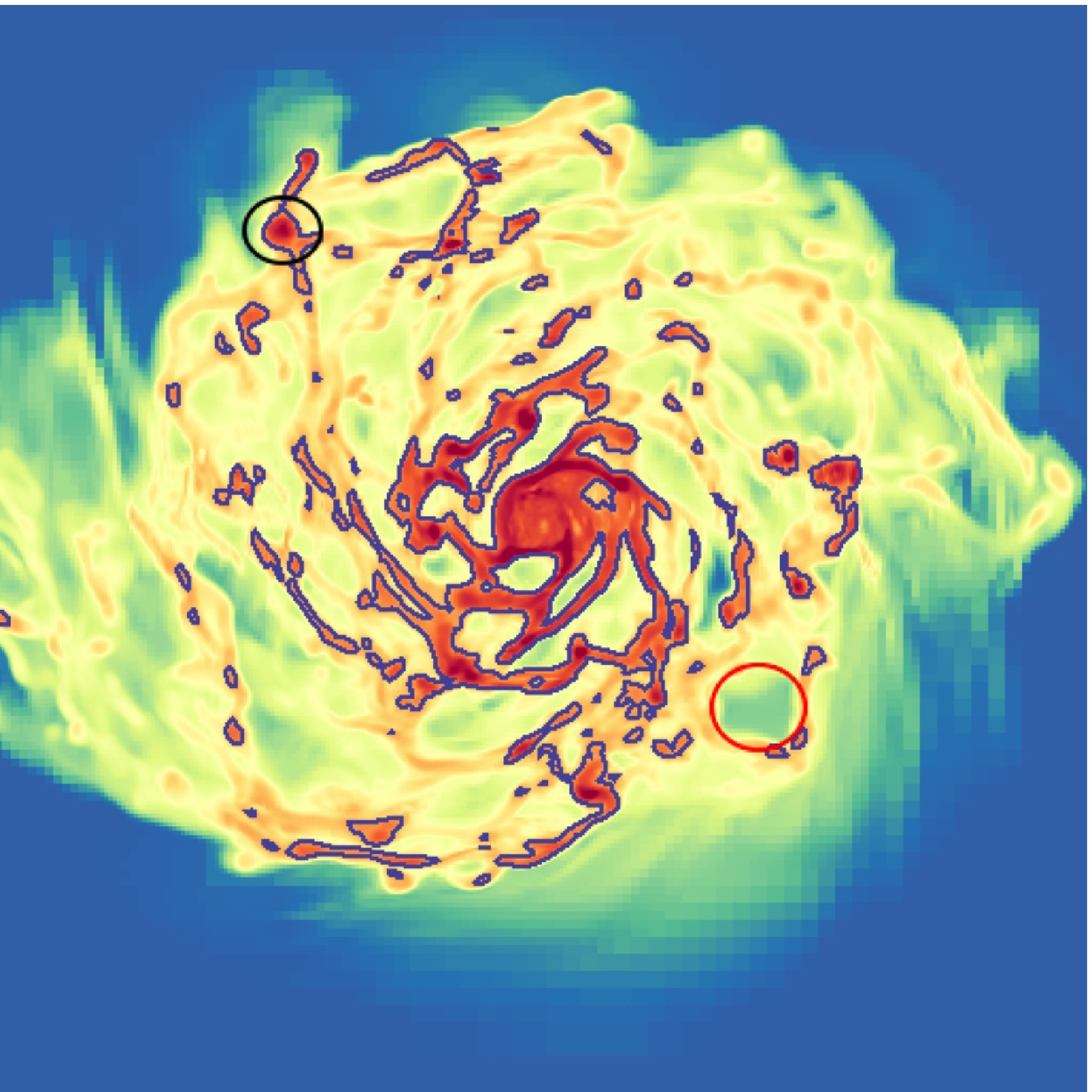}}}
  \centering{\resizebox*{!}{5.6cm}{\includegraphics{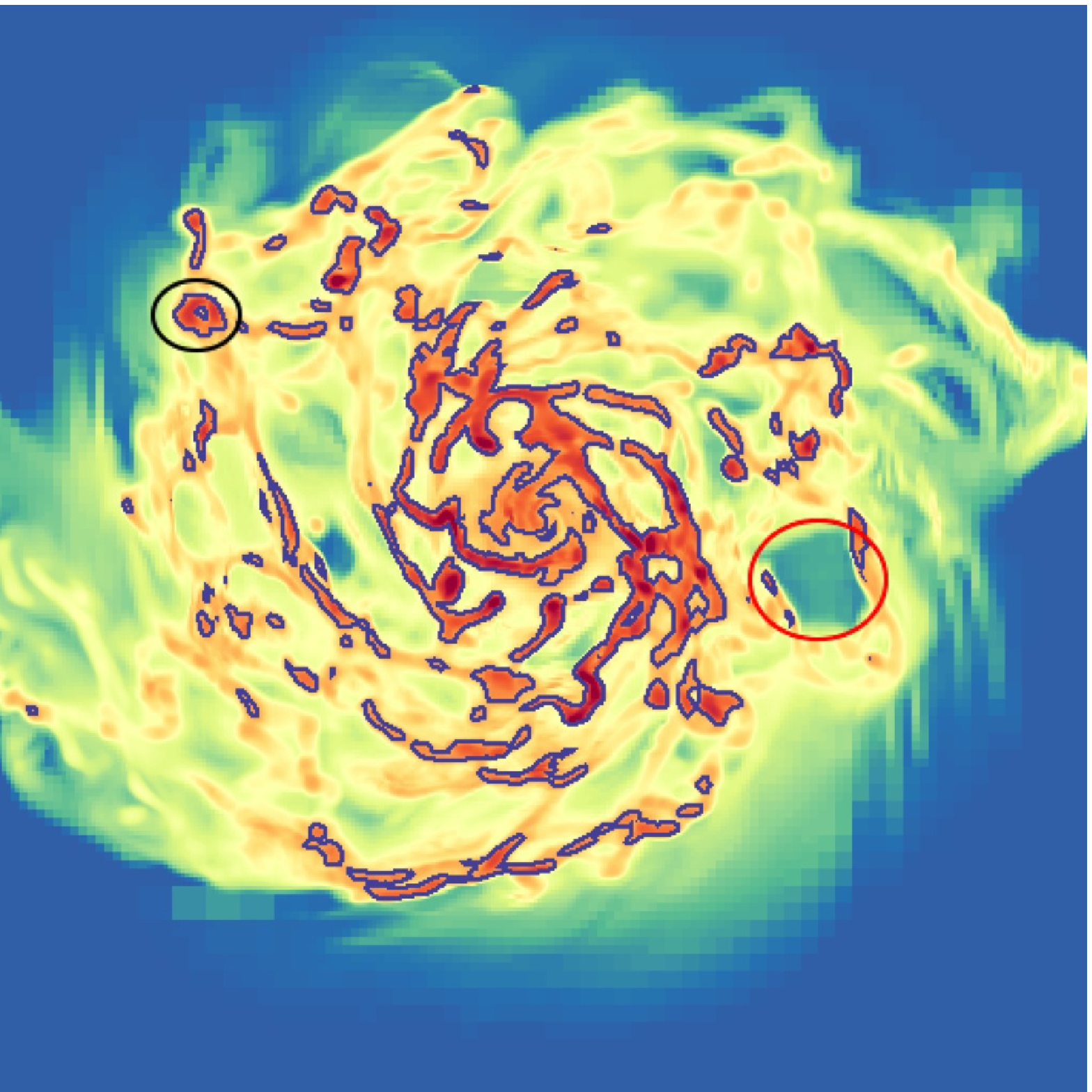}}}
  \centering{\resizebox*{!}{5.6cm}{\includegraphics{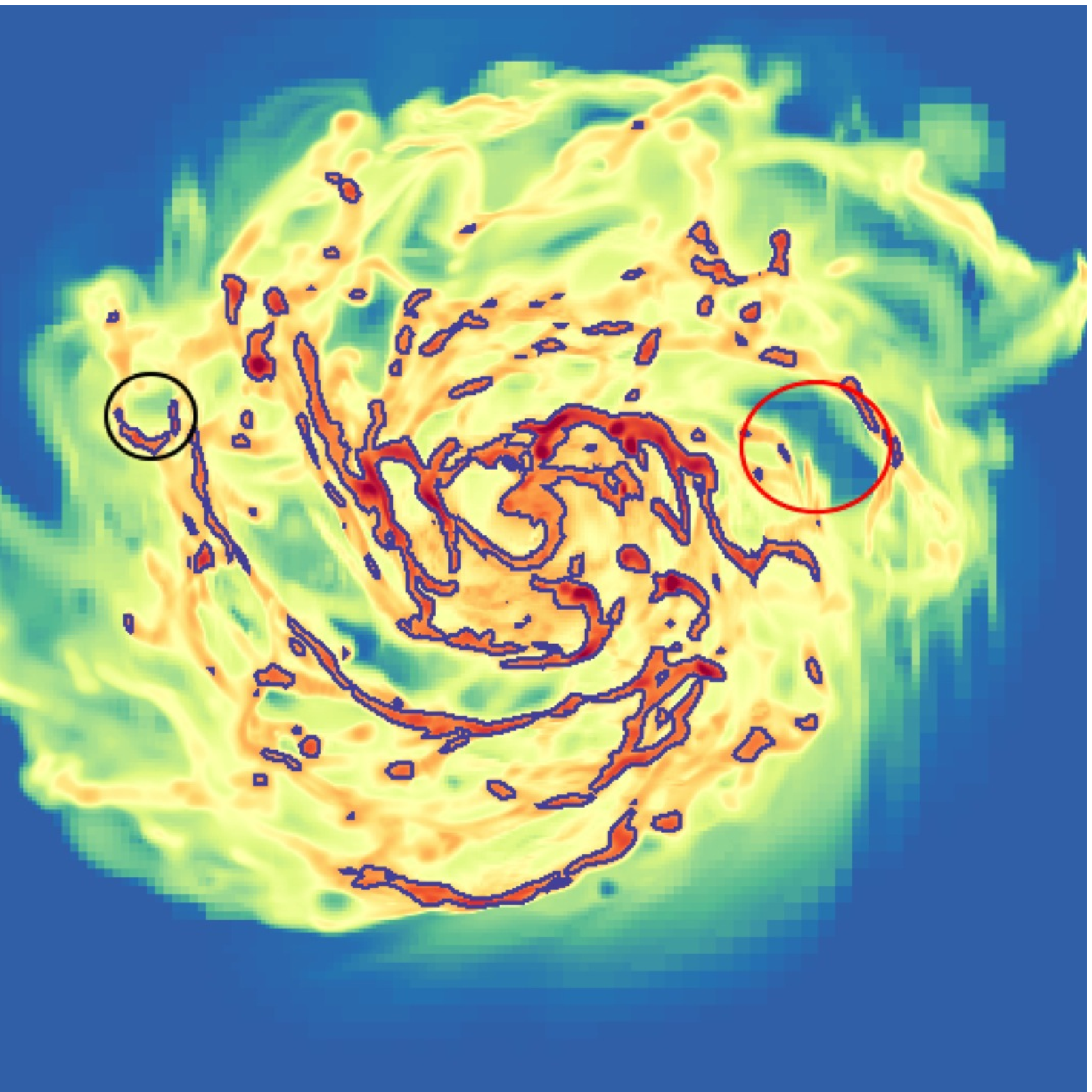}}}
  \caption{Isolated galaxy simulation. Time sequence of the gas density seen face-on at $t=610$~Myr from top left to bottom right, each snapshot is separated by 2 Myr. Image size is 5 kpc. This time sequence highlights the short-lived nature of small-scale substructures in the ISM. The colour table is from $0.1 \, \rm H\, cm^{-3}$ (blue) to $10^4 \, \rm H\, cm^{-3}$ (red). The blue contours surround the regions of star formation with gas density above $250 \, \rm H\, cm^{-3}$. Note the decrease in the gas density in the bulge of the galaxy. Two gas clumps are identified in the outskirts of the galaxy (black and red circles), which are destroyed by the SN explosions.}
    \label{fig:clouddisruption}
\end{figure*}

\section{Results}
\label{section:result}

\subsection{The isolated galaxy}

\subsubsection{Evaporation of the dense star-forming gas}

Fig.~\ref{fig:sfrisolated} shows the star formation rate (SFR) of the isolated galaxy over $1 \, \rm Gyr$ of evolution. 
The SFR is bursty with peaks of SFR larger than $50\, \rm M_\odot \, yr^{-1}$, and with an average value around $20-30 \, \rm M_\odot \, yr^{-1}$.
Note that the global behaviour of the SFR is to first increase until $\sim 500\, \rm Myr$ and decrease afterwards, which a generic result of the accretion rate on a galaxy from gas in a halo with an NFW profile~\citep{dubois&teyssier08winds}.
The large oscillations observed are due to the self-regulation of the cold star-forming gas in the ISM by SNe explosions~\citep{teyssieretal13}.
Here the peaks of the SFR are dominated by the activity within the bulge\footnote{Hereafter we denote indistinguishably the bulge and the central 500 pc region of the galaxy, which for an exponentially-decreasing spherically-symmetric profile with 200 pc scale radius (the value at $t=1$ Gyr) corresponds to the radius that contains 50 per cent of the mass.}, $r\le 500\, \rm pc$,
of the galaxy, while the average level of the SFR corresponds to the SFR in the disc of the galaxy within cold dense clumps.
During the beginning of the collapse of the halo $t \le 400\, \rm Myr$, several galactic outflows develop up to $5-10\, \rm kpc$ distance above the plane of the disc (see top panel of~Fig.~\ref{fig:outflow}).
However, due to the strong pressure confinement exerted by the ram-pressure of the hot infalling gas, the outflowing gas is maintained within that region and falls back onto the galaxy in a cycle of galactic fountain~\citep{dubois&teyssier08winds}.
Later on, after $t>400 \, \rm Myr$, when the activity of the SFR has decreased, small SNe bubbles rise a few hundred parsecs above the disc plane similar to what is observed in our own Milky Way disc (see bottom panel of~Fig.~\ref{fig:outflow}). 

In the disc, SFR and, thus, SNe feedback, are strongly clustered in small star-forming regions, where gas reaches densities above $250\, \rm H\, cm^{-3}$, and, in turn, strong localized feedback periodically disrupts the cold clumps of gas that are the nursery of stars.
This behaviour is illustrated in Fig.~\ref{fig:clouddisruption} with a time sequence of the gas density in the disc where two dense clumps of gas are pinpointed.
After a typical time of $t_{\rm SN}=10\, \rm Myr$ that corresponds to the average time for type II SNe to explode after the formation of the stellar particle, the whole clump is dispersed into the diffuse ISM.
Such clustered explosions naturally force the continuous driving of turbulence and the mixing of the various, hot/diffuse and cold/dense, phases of the ISM.
The same process happens in the bulge region where the star-forming gas is more extended.
The gas component is in the form of a compact disc exhibiting small spiral arms and after some time the whole region is dispersed by the heating of SNe feedback, increasing turbulence and therefore the overall velocity dispersion of the gas. 

We recall that the formation and rapid destruction of large-density clumps is the direct consequence of the adopted scheme for SNe feedback using delayed cooling.
This implementation is designed to affect the formation of such regions without radiating any of the thermal energy up to the point where the blast wave has propagated over a Jeans length.
Less efficient SN feedback implementations such as the one from~\cite{dubois&teyssier08winds} allow for the long term survival of such clumps, and the consequences on the central BH growth is investigated in detail in paper I.

\subsubsection{Evolution of the BH mass and spin}

Now that we have shown how SNe feedback alone is able to constantly reprocess the cold gas material in the ISM, we can investigate the growing mode of the central BH and in particular the evolution of its spin with respect to the accretion and gas dynamics.

\begin{figure}
  \centering{\resizebox*{!}{6.cm}{\includegraphics{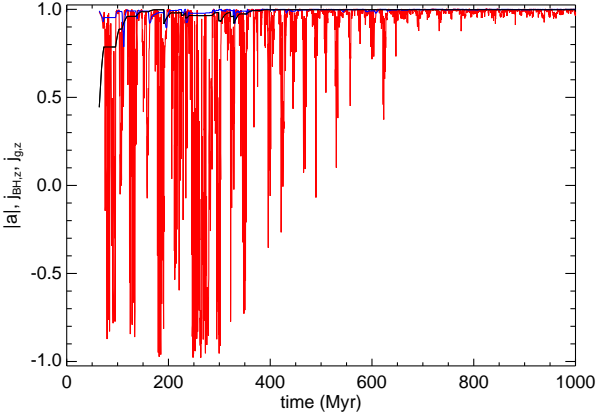}}}
  \caption{Isolated galaxy simulation. BH spin (black), vertical component of the BH AM (blue) and of the accreted gas AM (red) as a function of time over the whole run. The BH spin magnitude is always very large due to synchronisation between the BH spin direction and that of accreted gas AM during the phases of rapid growth, driven by cold, rotationally supported gas.}
    \label{fig:bhspin}
\end{figure}

\begin{figure}
  \centering{\resizebox*{!}{6.cm}{\includegraphics{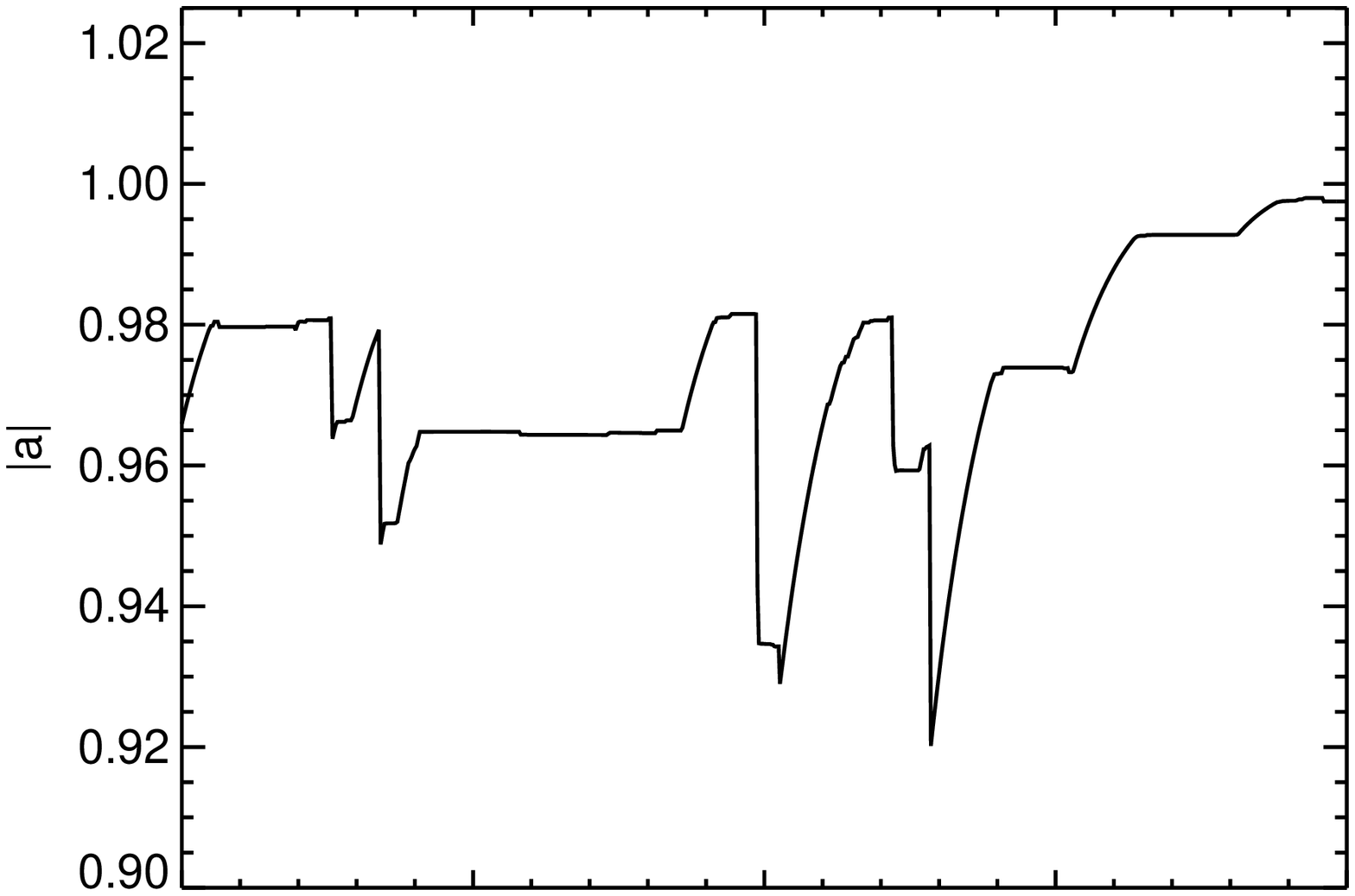}}\vspace{-1.35cm}}
  \centering{\resizebox*{!}{6.cm}{\includegraphics{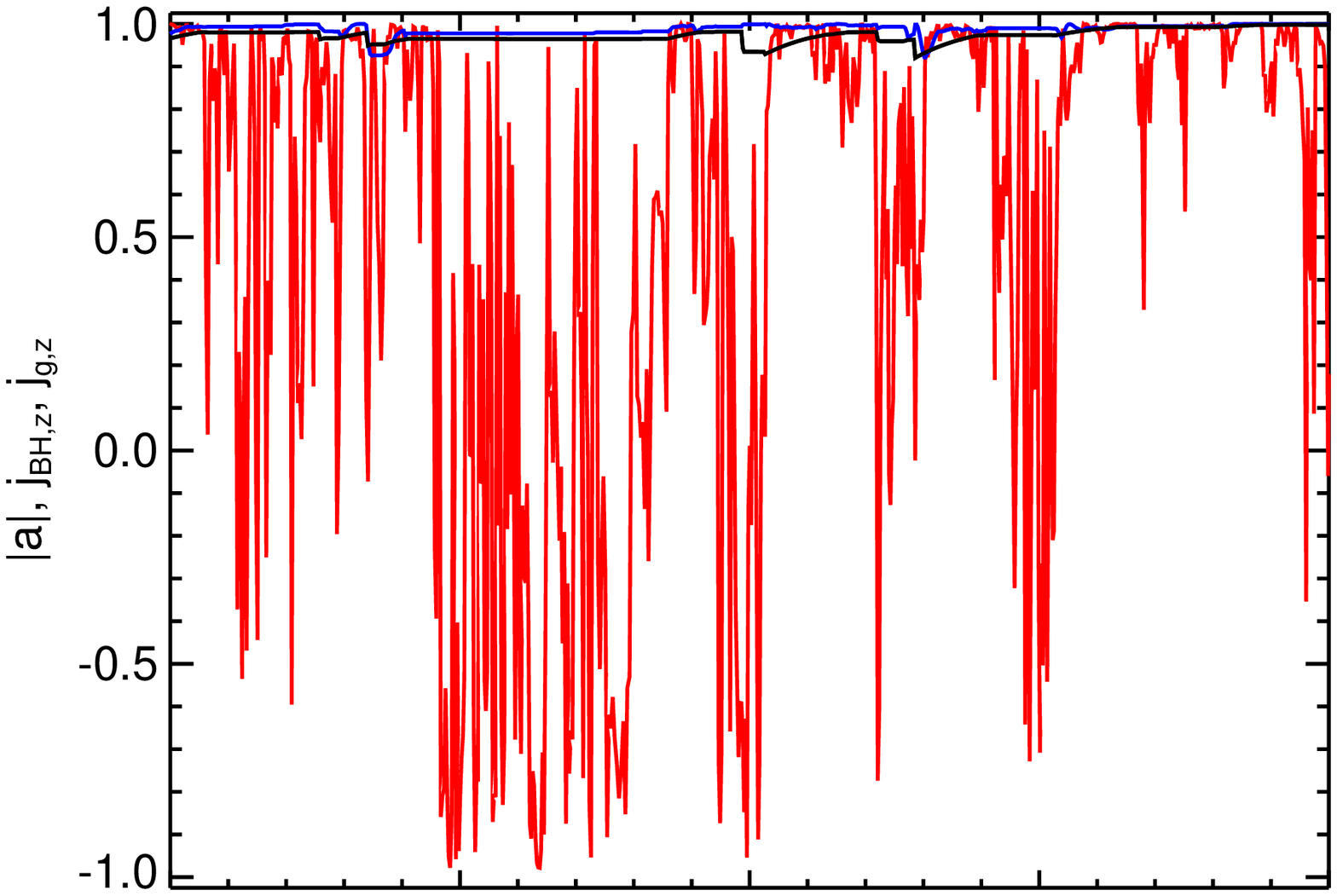}}\vspace{-1.35cm}}
  \centering{\resizebox*{!}{6.cm}{\includegraphics{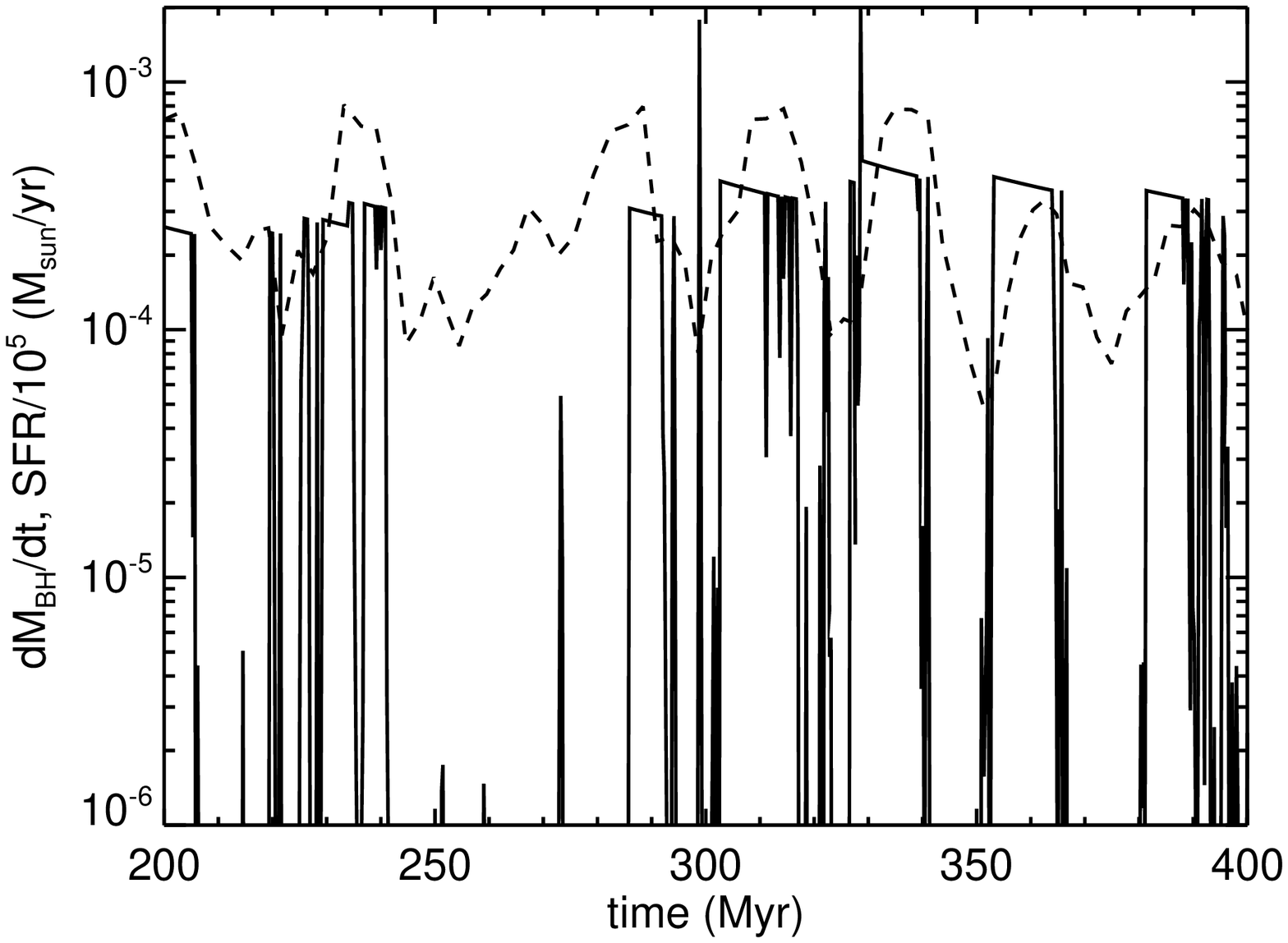}}}
  \caption{Isolated galaxy simulation. \emph{Top panel}: Spin as a function of time. \emph{Middle panel}: BH spin (black), vertical component of the BH AM (blue) and of the accreted gas AM (red) as a function of time. \emph{Bottom panel}: BH accretion rate (solid), SFR within the $r< 500 \, \rm pc$ bulge of the galaxy (dashed, in $10^5\, \rm M_\odot\, yr^{-1}$ units) as a function of time for 200 Myr of evolution, to highlight the short-term variability. }
    \label{fig:spinzoom}
\end{figure}

\begin{figure}
  \centering{\resizebox*{!}{6.cm}{\includegraphics{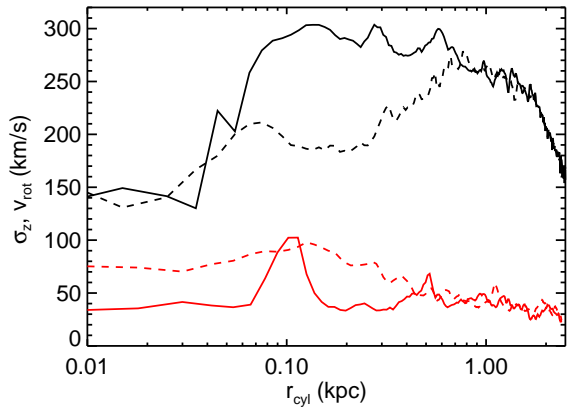}}}
  \caption{Rotational velocity (black) and vertical velocity dispersion (red) of the dense gas as a function of cylindrical radius at two different times $t=612\, \rm Myr$ when the BH accretes at Eddington (solid line) and at $t=620 \, \rm Myr$ when the accretion rate on the BH becomes negligible (dashed line). These times correspond to the top right and bottom right images of Fig.~\ref{fig:clouddisruption} respectively. Low accretion onto the BH happens when the velocity dispersion of the gas close to the BH increases. In phases of strong gas accretion, coherent gas rotation dominates over random motions in the gas.}
    \label{fig:sigmavsr}
\end{figure}

Fig.~\ref{fig:sfrisolated} shows the evolution of the BH mass with time. 
The BH starts with an initial seed mass of $10^4\, \rm M_\odot$ and grows up to $M_{\rm BH}\simeq 3\times 10^5\, \rm M_\odot$ after $1 \, \rm Gyr$.
There is no self-regulation (except after $t_{\rm end} > 800$ Myr), thus the BH growth is limited at the Eddington rate. However as the spin magnitude is of the order of 1, the radiative efficiency is $\epsilon_{\rm r}\simeq0.32$ instead of 0.1 as usually assumed from the Soltan argument, and the Eddington accretion rate is decreased by a factor 3. 
Also the BH grows in an episodic fashion due to SN explosions with periods limited by Eddington and periods with almost zero gas accretion: both periods having a   similar time duration. 
Therefore, we can simply multiply the Salpeter ($t_{\rm Salp}=\epsilon_r \sigma_T c/(4\pi G\,m_p)=0.045 (\epsilon_r/0.1)\,{\rm Gyr}$) time-scale by a factor 2 when comparing the analytical growth to the simulated one. 
The power law behaviour can be readily understood as $M(t)=M_0 \exp(t/\tau)$ where $\tau=2\times t_{\rm salp}$, scaled appropriately for the problem, i.e., in this case increased by a factor of three because of the radiative efficiency ($\epsilon_r=0.32$), and multiplied by a factor of two because of the intermittent accretion, effectively halving the time a BH is accreting.

The spin evolution of the BH in Fig.~\ref{fig:bhspin} does not show any large variation.
The value of the spin is always above $|a|>0.9$ after the first e-folding time-scale of the BH growth.
We see some variations of the order of $\delta a=0.1$ but, soon after,  accretion is strong enough to rapidly re-align gas and BH spin, and bring the spin of the BH back to $|a|>0.9$.
Recall that if  accretion is completely random in terms of AM direction (and constant in accretion rate), the spin of the BH would be close to zero as a BH gets spun down faster than it is spun up. Conversely, if  accretion is coherent in AM direction, or at least has some  level of coherence~\citep{dottietal13}, the BH keeps very high spin values.

The reason for this large value of the BH spin has to do with both the orientation of the gas in the surroundings and the synchronisation with events of strong accretion driven by cold and dense gas.
The blue curve of Fig.~\ref{fig:bhspin} shows that the direction of the BH AM is always pointing towards the vertical direction and is, thus, well aligned with the galactic AM.
The red curve shows instead the z-direction of the AM of the gas accreted onto the BH (red curve in Fig.~\ref{fig:bhspin}). During the galactic fountain phase ($<400$ Myr), this direction can be rapidly changing, while at a later stage, when the galactic fountain is replaced by low-rising bubbles, the gas AM realigns with the vertical direction of the disc.

If the direction of the accreted gas AM is quickly changing in the galactic fountain phase, why is the BH spin unaffected? 
A careful inspection of such ``chaotic'' events is shown in Fig.~\ref{fig:spinzoom}. Some time intervals with opposite accreted gas AM and BH AM are clearly identified, such as at $245<t<285\, \rm Myr$, and should, in principle, decrease the BH spin. However, this prolonged chaotic period is also synchronised with a period of extremely low accretion rate onto the BH, $\dot M_{\rm BH} \ll \dot M_{\rm Edd}$, so that the BH does not accrete enough gas to decrease durably and significantly its spin amplitude.
\emph{Vice versa}, prolonged high-accretion rates onto the BH, $\dot M_{\rm BH}\sim \dot M_{\rm Edd}$, correspond to accreted gas AM aligned with both the BH AM and the galactic disc (with the vertical axis). The sustained BH accretion events are driven by the replenishing of the cold, dense gas reservoir in the bulge of the galactic disc, which is also synchronised with the peak of the SFR, and in that case the whole central region co-rotates with the galactic disc.

The BH grows for the most part accreting smooth gas rather than clumps. The accretion of clumps is  limited because these clumps have extremely short life-times ($\sim 10 \, \rm Myr$) and most of them can not reach the bulge before being destroyed by SNe.
In the case where clumps are more long-lived, the situation could be different.
This alternative scenario has been tested in Appendix A of Dubois et al. (2013), and is reproduced in the Appendix~\ref{appendix:snkinetic} of the current paper, with inefficient SNe feedback (long-lived star-forming clouds), and we found that even though the BH can grow by the capture of gas clumps (especially at late times), the gas AM of the accreted clumps is typically well-aligned with that of the underlying gas disc. 

In Fig.~\ref{fig:sigmavsr} we show the vertical velocity dispersion of the dense gas with $n > 1 \, \rm H\, cm^{-3}$ at two different times when the BH accretes at Eddington corresponding to the top right panel of Fig.~\ref{fig:clouddisruption}, and when the accretion drops below $10^{-3} \dot M_{\rm Edd}$ corresponding to the bottom right panel.
The gas velocity dispersion and rotational velocity in the central region, very close to the BH, are of comparable value, with $150\, \rm km\, s^{-1}$ for the rotational velocity, and $40$ and $70\,Ê\rm km\, s^{-1}$ for the velocity dispersion respectively before and while the gas is blown away from the centre due to SNe energy release.
Outside the very centre, the gas rotational velocity increases up to $v_{\rm rot}\simeq250\, \rm km\, s^{-1}$ and the gas velocity dispersion goes to $\sigma_{\rm z}\simeq50\, \rm km\, s^{-1}$.
We also clearly see the effect of the SNe energy release in the velocity patterns, both in the rotation of the gas and in the velocity dispersion in the $50<r_{\rm cyl}<500\, \rm pc$ radius range: SNe explosions temporarily increase the velocity dispersion of the gas, while they reduce the velocity rotation.
The fact that the gas velocity dispersion does not overwhelm the gas motion near the BH explains why the BH spin keeps very large values and is consistently pointing towards the same direction, which is the one of the gas AM around the BH.

\subsection{The {\sc Seth} simulation: a cosmological laboratory for BH spin evolution}

\begin{figure}
  \centering{\resizebox*{!}{6.cm}{\includegraphics{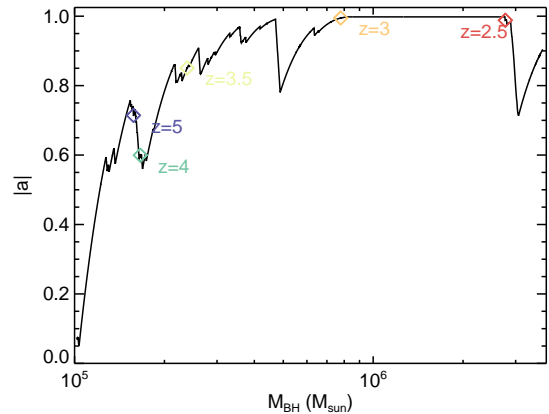}}}
   \caption{{\sc Seth} cosmological zoom simulation.  Spin magnitude versus mass for the BH of the central most massive galaxy. Different values of redshift are indicated with diamonds. The spin grows to its maximum value at $z\sim3$, and then remains above $0.7$.}
    \label{fig:mbhevol_z7-3}
\end{figure}

\begin{figure}
  \centering{\resizebox*{!}{6.cm}{\includegraphics{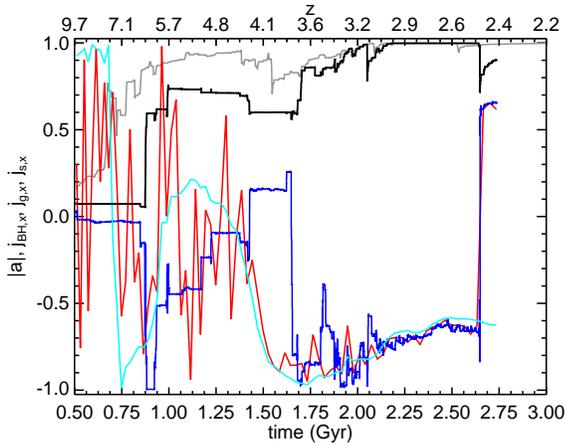}}}
  \caption{{\sc Seth} cosmological zoom simulation.  Magnitude of the BH spin (black), x-component of the gas AM within 500 pc (red) of the BH spin (blue) and of the stars within the galaxy (cyan) as a function of time and  redshift. The light grey line is for the {\sc Seth} low-resolution simulation with $\Delta x=40 \, \rm pc$. At early stages ($\sim 1.5 \, \rm Gyr$), the gas behaviour is chaotic and shows large fluctuations of the AM direction. At later times, gas settles in the galaxy and there is a strong persistent correlation between the BH spin, the gas AM  and that of the stars in the galaxy, which leads to relatively high spin values.}
    \label{fig:abhevol_z7-3}
\end{figure}

\begin{figure*}
  \centering{\resizebox*{!}{5.6cm}{\includegraphics{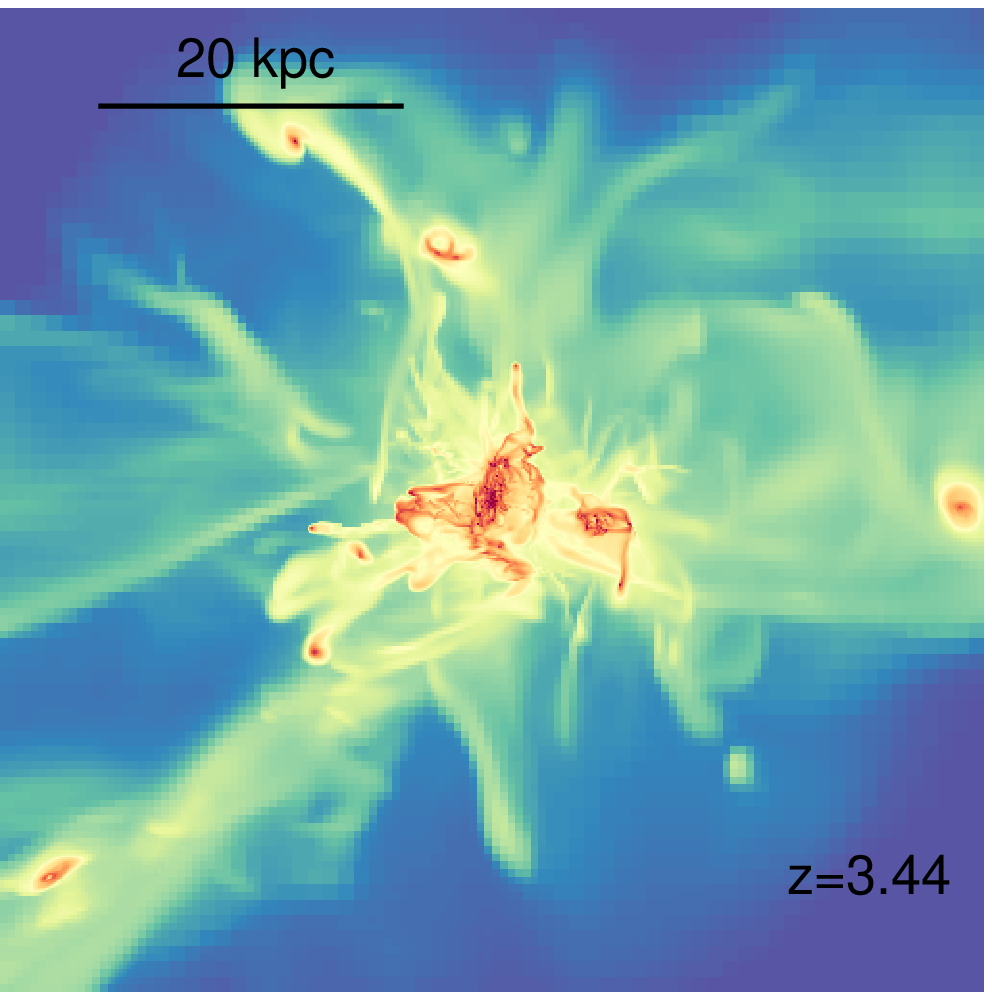}}}
  \centering{\resizebox*{!}{5.6cm}{\includegraphics{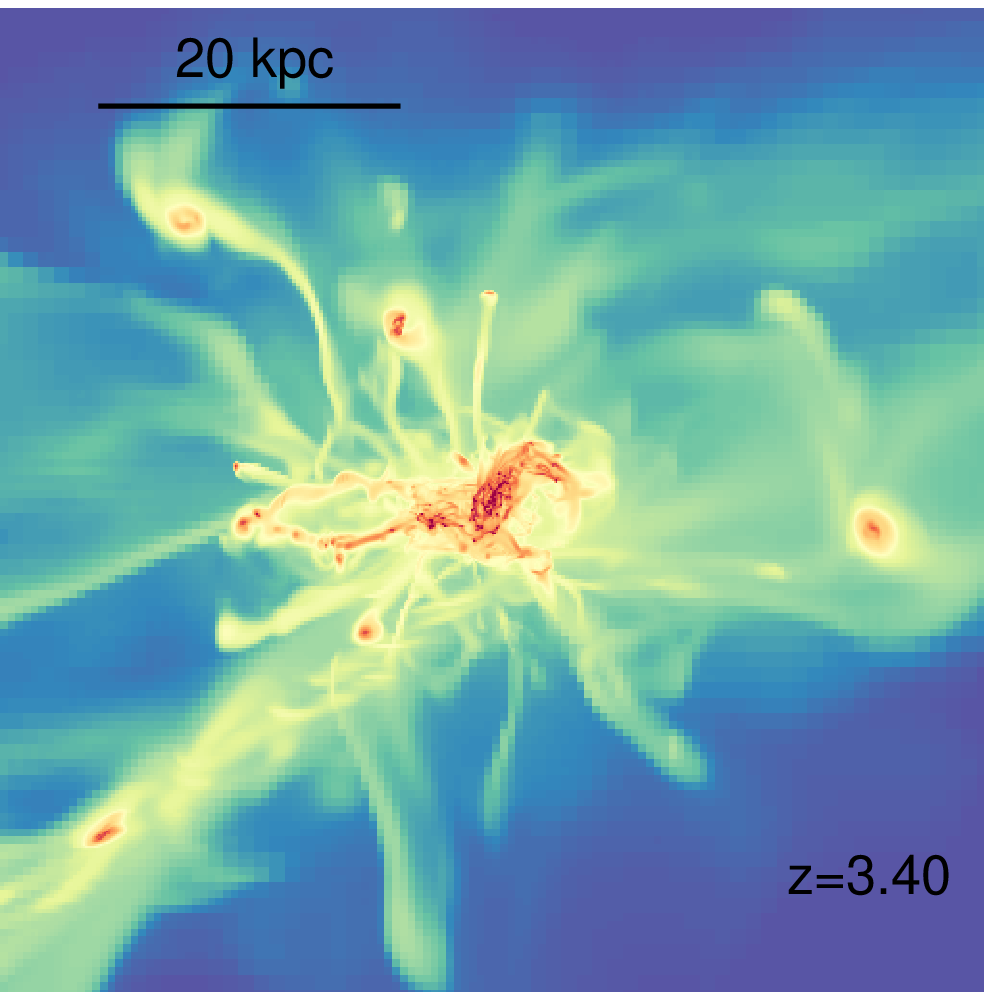}}}
  \centering{\resizebox*{!}{5.6cm}{\includegraphics{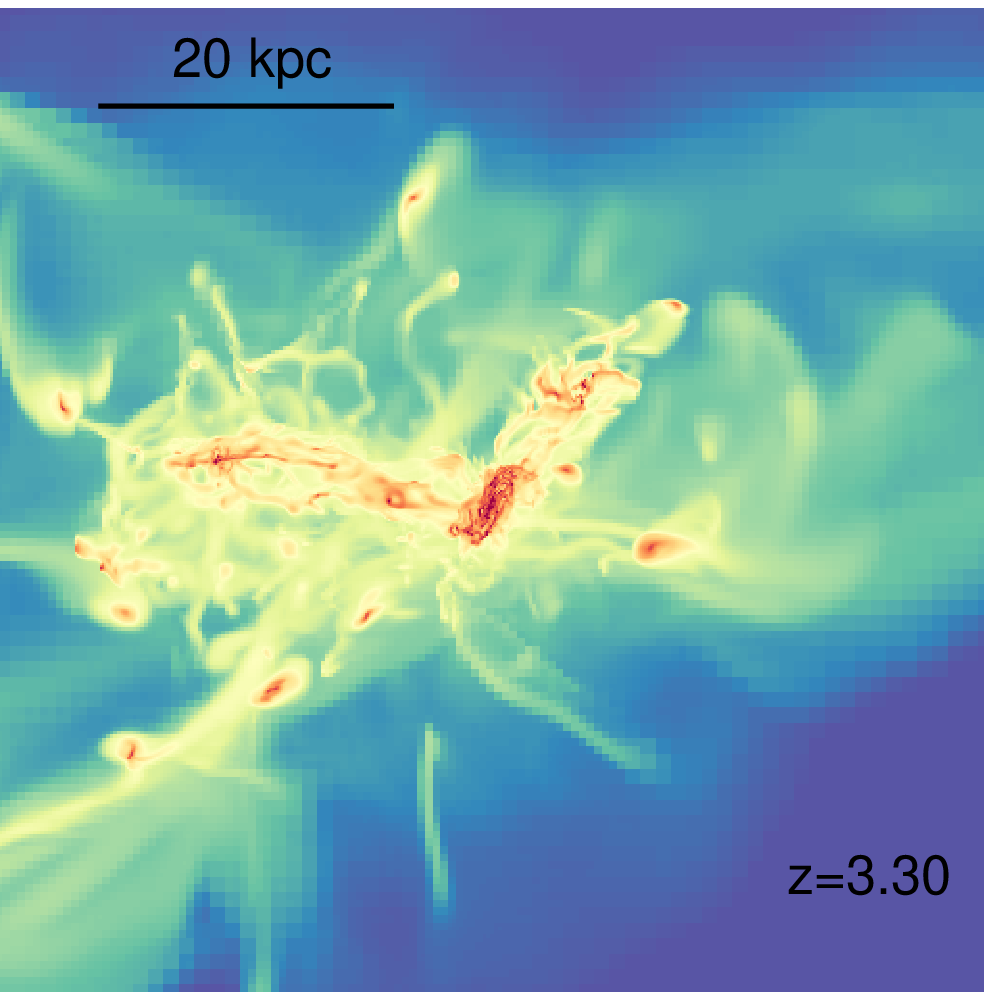}}}
  \centering{\resizebox*{!}{5.6cm}{\includegraphics{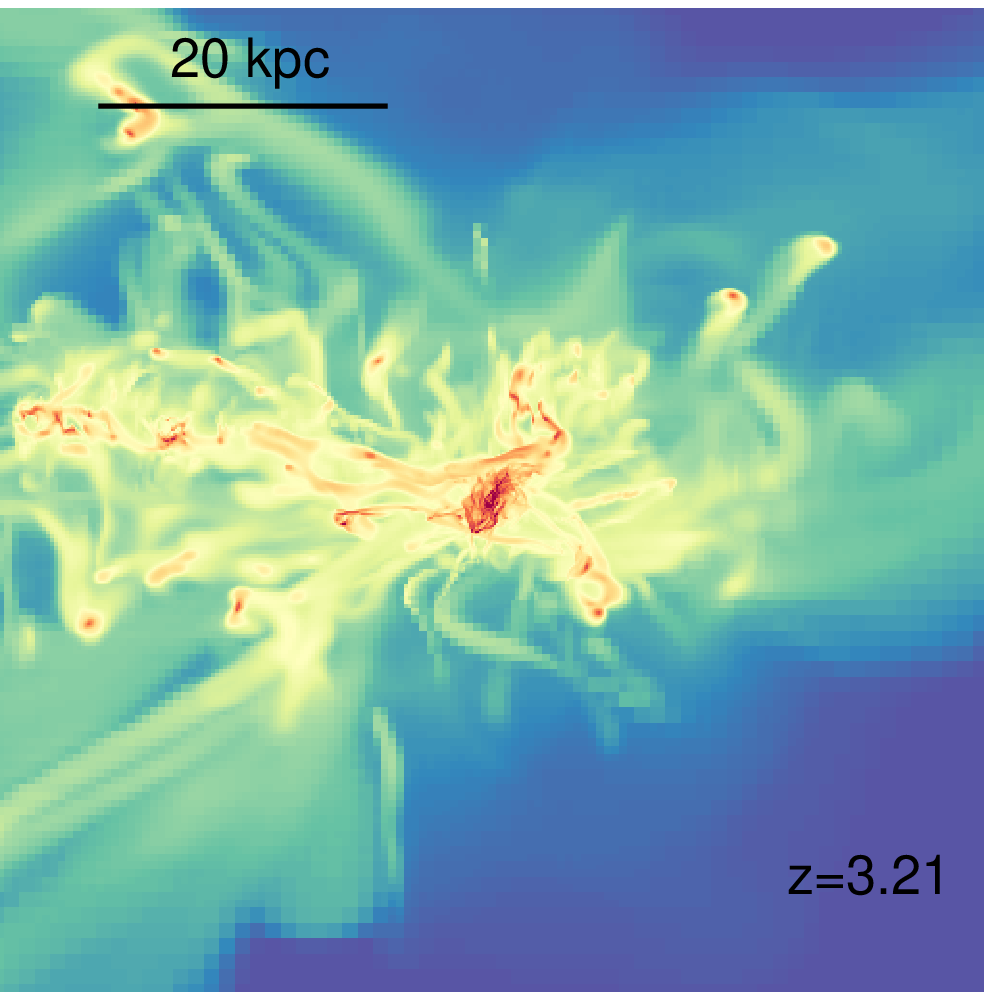}}}
  \centering{\resizebox*{!}{5.6cm}{\includegraphics{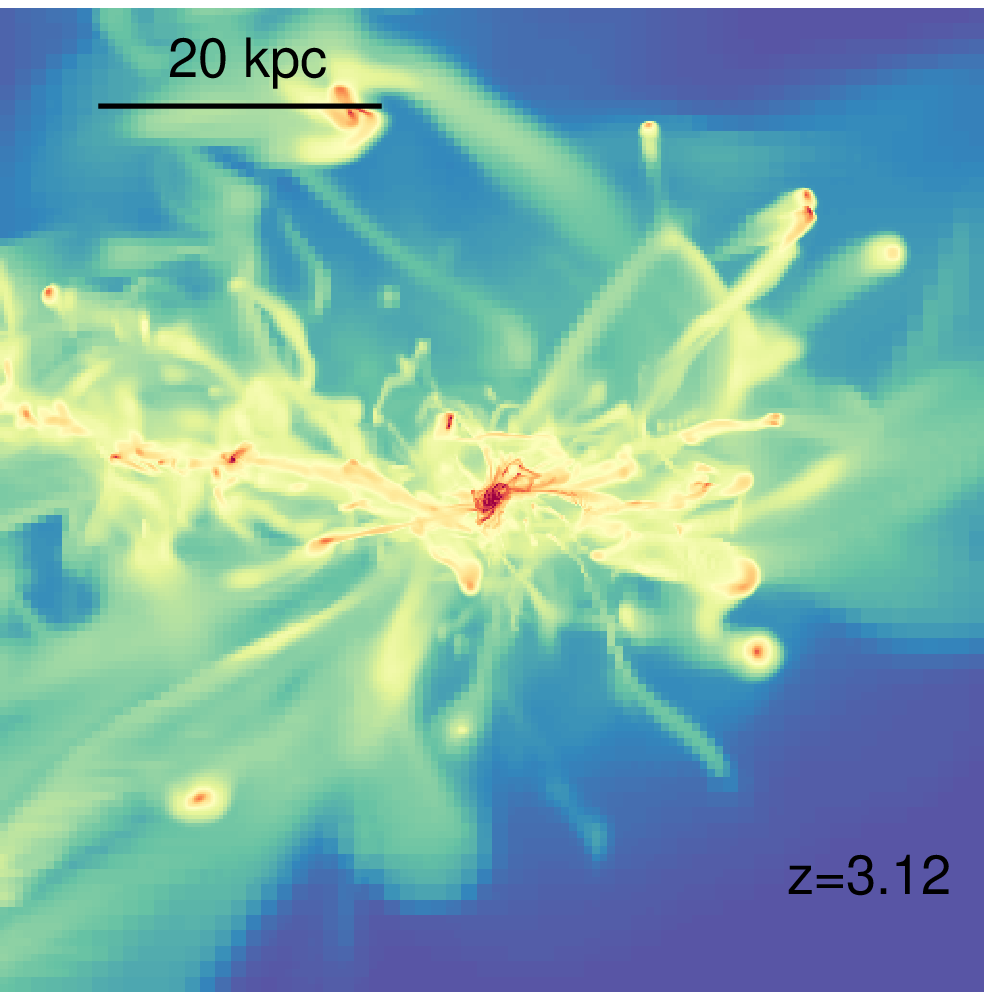}}}
  \centering{\resizebox*{!}{5.6cm}{\includegraphics{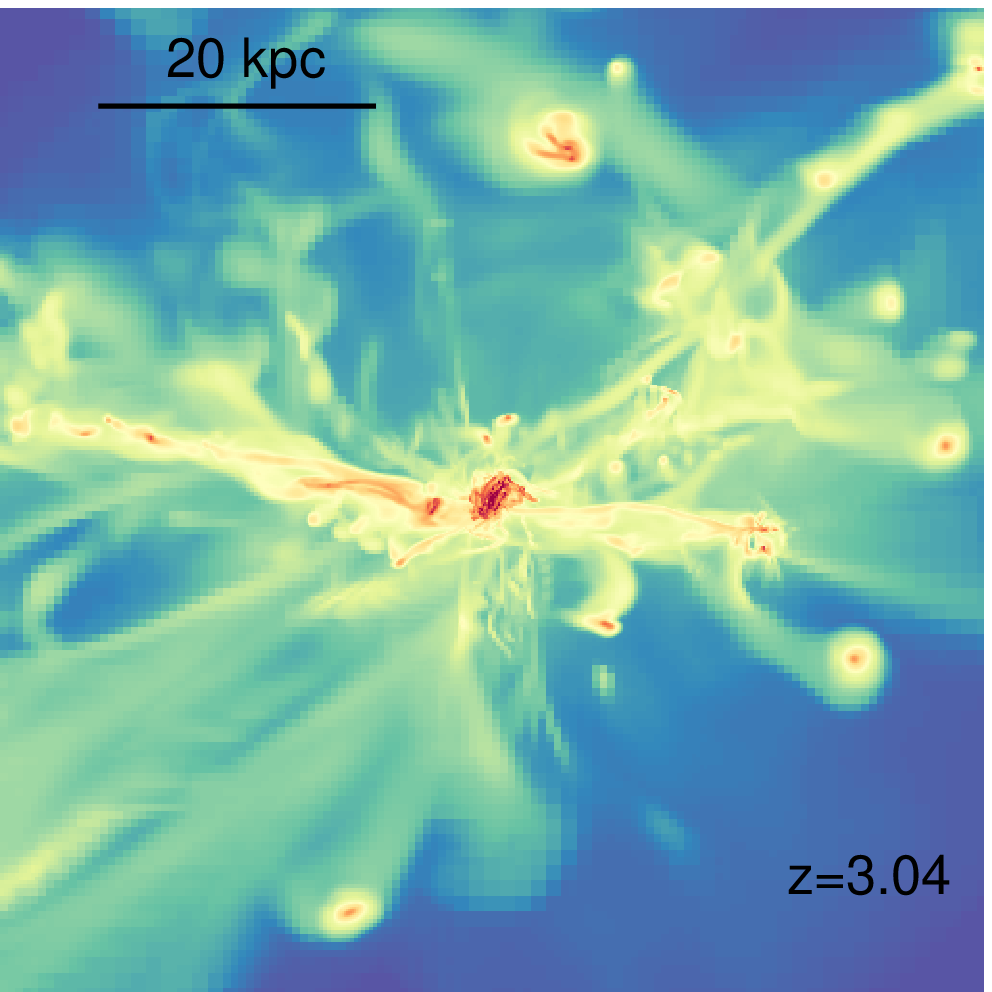}}}
  \caption{{\sc Seth} cosmological zoom simulation. Projected gas density at different redshifts illustrating the merger of one gas-rich satellite with the central galaxy. The colour table is from $10^{-3} \, \rm  H\, cm^{-3}$ (blue) to $10^{2} \, \rm  H\, cm^{-3}$ (red). The virial radius of the underlying dark matter halo is $\simeq 50\, \rm kpc$. Some dense filamentary gas is stripped out during the head-on collision and falls back onto the galaxy. }
    \label{fig:merger}
\end{figure*}

\begin{figure}
  \centering{\resizebox*{!}{6.cm}{\includegraphics{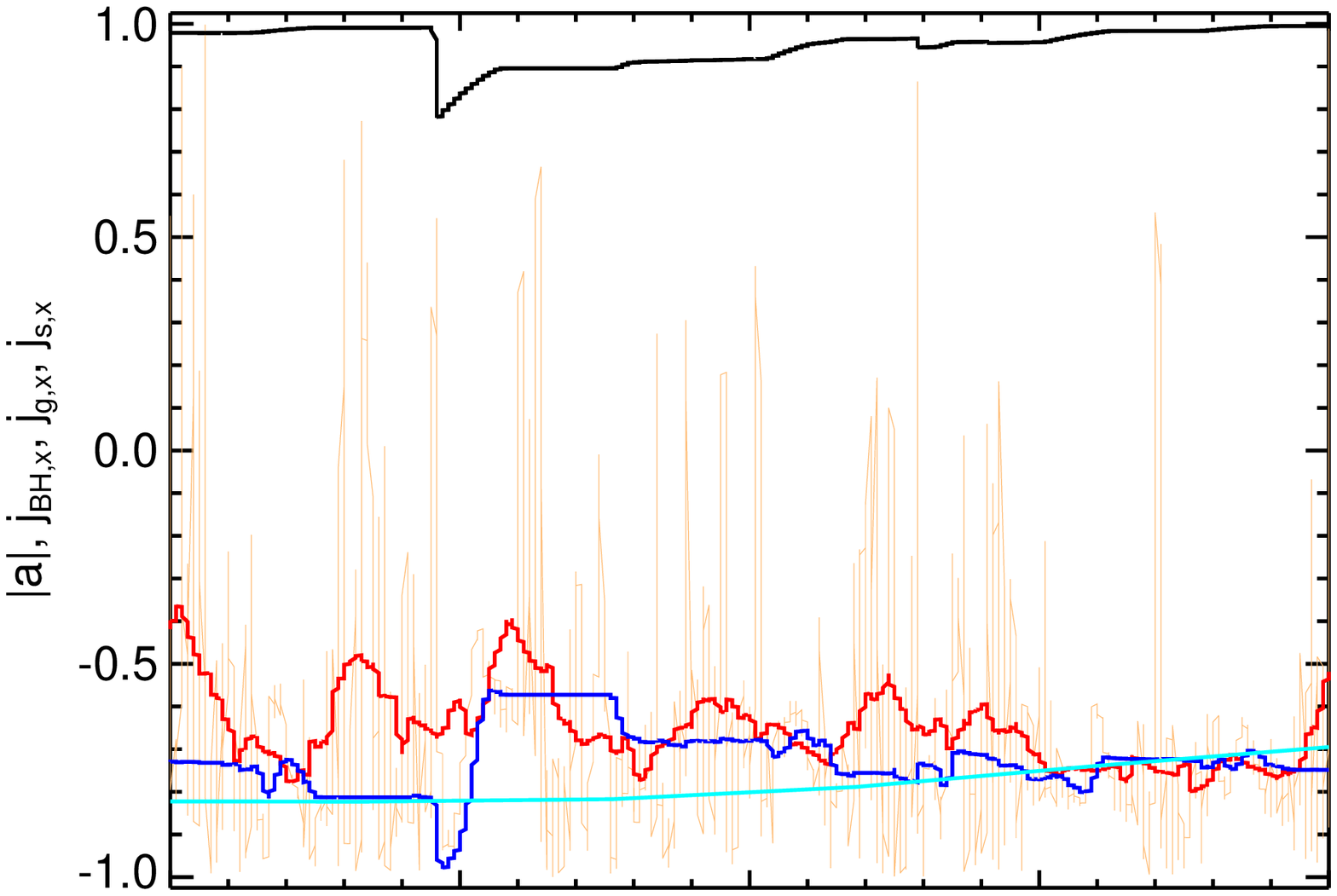}}\vspace{-1.35cm}}
  \centering{\resizebox*{!}{6.cm}{\includegraphics{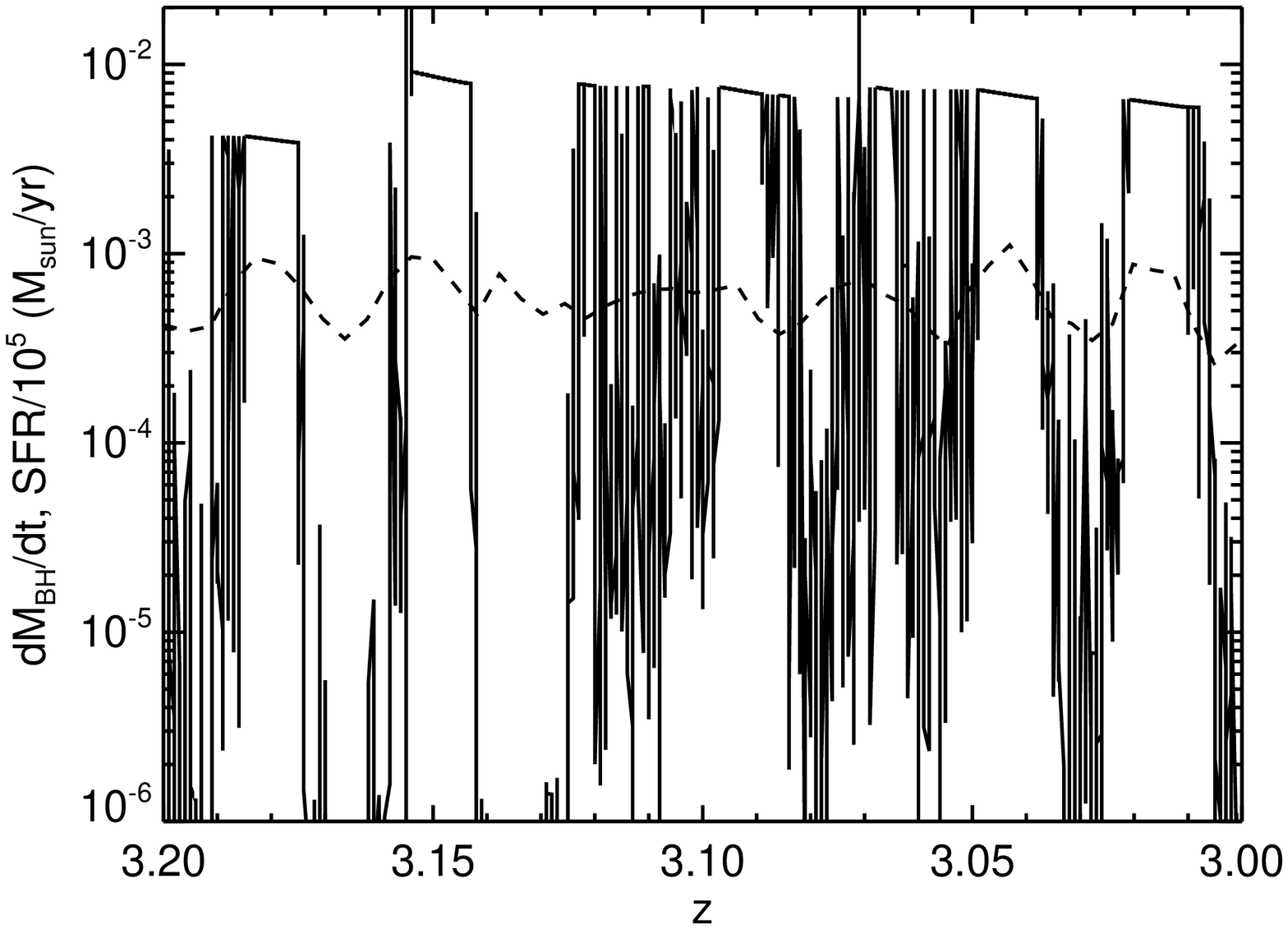}}}
  \caption{{\sc Seth} cosmological zoom simulation.  
  \emph{Top panel}: Absolute value of the BH spin versus redshift (black), x-component of the accreted gas AM (instantaneous in orange and smoothed in red), of the spin (blue) and of the stars within the galaxy (cyan). \emph{Bottom panel}: BH mass accretion rate (solid) and SFR$/10^5\, \rm M_\odot$ (dashed). Eddington-limited accretion episodes are synchronous with a strong correlation between the directions of the accreted gas AM, the BH spin and the galactic AM.}
    \label{fig:mbhevol}
\end{figure}

As discussed in Paper I, the resolution employed in {\sc Seth} allows us to resolve the multiphase structure of the galaxy with star-forming clumps of dense cold gas, hotter diffuse gas. At relatively late cosmic times, $z<3.5$, it also allows us to identify the formation of a razor thin disc of stars, a hotter and thicker disc of stars, and old clusters of stars flying around the galactic disc. 
At earlier epochs though, the galaxy's disc has not settled, as the gas is constantly stirred and reprocessed by the efficient feedback from SNe. The gravitational potential of the galaxy and of the host halo is not deep enough to prevent the galactic wind from escaping and gas rarely condenses in the central region close to the BH.  Above $z>3.5$ the growth of the BH is very slow, as shown in Fig.~\ref{fig:mbhevol_z7-3}: BH growth is delayed and the mass of the BH only increases by a factor 2. When galaxy and halo have grown sufficiently, their gravitational potential is able to retain the gas affected by SN feedback and a gaseous disc finally forms (around $z\sim 3.5$). Cold gas starts to accumulate within the central region around the BH. Since this gas originates from the recently settled disc, the AM of the gas that feeds the BH tends to have a coherent AM direction. 

Fig.~\ref{fig:abhevol_z7-3} shows that above $z>3.5$ (before $t<1.8  \, \rm Gyr$) the magnitude of the BH spin is initially constant over long timescales, while the accretion rate on the BH is small, and spin changes only when a gas clump approaches. In this phase the gas AM in the central region (500 pc) is randomly oriented because SNe explosions easily eject gas out of the galaxy, precluding the settling of a persistent gas disc.  The accreted gas AM - not represented in Fig.~\ref{fig:abhevol_z7-3} - shows an even more chaotic behaviour early on, with sudden variations from one coarse time step to another. Once the gas in the galaxy starts to settle in a rotationally supported disc (below $z<3.5$), because SNe become unable to eject gas out of the galaxy, the BH spin starts increasing, until it reaches its maximum magnitude at $z\sim 3$. Meanwhile, the accreted gas AM becomes coherent in time and well aligned with the BH spin direction and that of the AM of the galaxy (defined here as the AM of the stars).

Several external perturbations caused by satellites flying around the halo are observed at any redshift. There is one particular event where a gas-rich merger produces a large ejection of gas from the central galaxy into the circum-galactic medium (Fig.~\ref{fig:merger}). A gas-rich satellite galaxy at $z=3.44$ collides with the central one head-on, passing through the galactic disc and stripping gas. When this cold (a few $10^4\, \rm K$) moderately-dense (a few $0.1\, \rm H\, cm^{-3}$) gas falls back onto the galaxy, it is misaligned with the original gas AM. When this gas reaches the BH, it triggers accretion of counter-rotating material and the spin magnitude decreases as shown in Fig.~\ref{fig:mbhevol} at $z=3.15$.

Towards the end of the simulation, $t=2.65 \, \rm Gyr$ ($z=2.5$), the accreted gas AM suddenly counter-rotates with respect to both the BH spin and stellar AM. This is caused by a large-scale cold gas filament which swaps sides in its connection to the galaxy, all of a sudden providing its AM in the opposite direction ~\citep[an effect already outlined in][]{tillsonetal12}.
This event decreases the spin amplitude, but the direction of the BH spin eventually re-aligns with the direction of the gas AM before the spin down process is completed~\citep{dottietal13}.
Note that the result is not extremely sensitive to a change in spatial resolution: for the {\sc Seth} simulation with $\Delta x=40 \, \rm pc$ resolution the BH spin grows slowly to large values at early times, and keeps a value close to 1 when the disc is settled, while gas-rich mergers and counter-rotating cosmic filamentary accretion are decreasing the value of the spin at almost the same positions in time (see Fig.~\ref{fig:abhevol_z7-3}).

As for the isolated galaxy simulation, the BH growth is episodic with short bursts of accretion at or close to the Eddington limit, followed by periods of negligible accretion ($\dot M_{\rm BH}<10^{-3}\, \rm M_\odot$) as we can observe in Fig.~\ref{fig:mbhevol}. The Eddington-limited accretion events are synchronous with the peaks of the star formation rate in the galaxy and, in particular, with the peaks of the SFR in the bulge, when cold dense gas is present around the BH.

The same figure (Fig.~\ref{fig:mbhevol}) also displays the direction of the BH spin, accreted gas and galaxy AM with respect to a fixed reference frame.
They are all very coherent over time, all pointing towards the same direction and with very little variation.
The direction of the accreted gas AM is {\it on average} pointing in the same direction during long extended period of times (red curve), while it {\it instantaneously} swings because of the efficient feedback produced by local SN explosions. However, since these explosions inject energy in the cold, dense gas, transforming it into diffuse and hot material,  accretion onto the BH drops to extremely low values and these chaotic periods cannot effectively spin down the BH. Note also the slight drift of the galaxy AM (stellar component) over time which is the result of smooth torques exerted by the external perturbations (cold streams, diffuse accretion of gas, minor mergers), while at some very specific moments the spin of the galaxy changes due to more significant mergers (e.g. at $z=5.9$ and $z=4.3$, see Fig.~\ref{fig:abhevol_z7-3}).

The direction of jets is not always observed to be aligned with the galaxy plane and can exhibit sharp bends (see~\citealp{BattyeBrowne2009} and~\citealp{BrowneBattye2010}).
This is in apparent contradiction with our simulations where BH spins tend to align with that of the galaxy at high redshift.
However, the orientation of jets is not necessarily aligned with BH spins because the multiphase structure of the ISM can eventually jerk the jet on larger scales. 
Moreover, mergers in gas-poor galaxies are responsible for misaligning the orientation of the BH spin with that of the host galaxy (see paper III).

\section{Caveats}

The Bondi-Hoyle radius $R_{\rm Bondi}=GM_{\rm BH}/c_{\rm s}^2$ is under certain circumstances resolved in our simulation. 
For a typical $M_{\rm BH}=10^6 \,Ê\rm M_\odot$ BH in a gas with temperature $10^4 \, \rm K$, the corresponding Bondi radius is 40 pc (4 pc if $M_{\rm BH}=10^5 \,Ê\rm M_\odot$). 
Therefore, when the BH is in dense and cold gas, the Bondi radius is resolved and leads to Eddington-limited phases.
However, when the gas temperature increases because of a SN feedback episode or due to central BH feedback, the Bondi radius becomes under-resolved.
The latter corresponds to phases with low accretion rate, probably with a value which is not exactly the one that we might obtain by resolving the Bondi radius, though the accretion ration would remain low (much lower than Eddington) because of the removal of the dense gas by feedback.

The sub-grid model for the feedback from the central AGN assumes that the BH produces thermal blast waves in the quasar mode. 
It is a crude approximation to the real feedback from a thin disc accretion disc, which should produce photons and a spherical wind that would propagate to larger-scales.
Detailed simulations have shown that the accretion on the central BH becomes episodic once radiation transport in dense environment is included~\citep{milosavljevicetal09, park&ricotti11, park&ricotti12}. 
However, these simulations assume a uniform background, and a more realistic geometrical configuration of the flow would probably help the photons to propagate along the path of least resistance without affecting the bulk of the gas accretion.

Though our resolution tests do not indicate that the AM on small-scales ($\sim 10\, \rm pc$) violently de-correlates with that on large galaxy scales ($\gtrsim 100\, \rm pc$) as we increase the resolution, there is still some unresolved turbulence down to the disc gravitational radius that remains to be resolved.
The gas continues to lose AM below $\sim 10$ pc scales~\cite[e.g.][]{wiseetal08, reaganetal14}, but it is still unclear wether this loss of amplitude goes along with a loss of coherence of the direction of the flow and if such a loss of coherence is maintained for sufficiently long times.

\section{Conclusions and Discussion}
\label{section:conclusion}

By means of high-resolution hydrodynamical simulations, we have followed the SN driven reprocessing of cold star-forming gas into a hot diffuse interstellar medium and its effect on the evolution of the spin of central supermassive BH.  All these physical quantities were evolved on-the-fly in a self-consistent manner. 
The turbulence induced by stellar feedback is described by  `maximal' model, to highlight its impact on the AM of the gas accreted by the BH. 
With two different kinds of initial conditions: an isolated collapsing halo and a zoomed cosmological halo evolved for several hundreds of Myrs, we reach similar conclusions. SN-driven turbulence does not contribute to randomising the AM of the gas effectively accreted by the BH. This is because the BH grows in mass by accreting cold, dense gas, which tends to reside in a rotationally supported disc. The gas made turbulent by SN energy injection is instead hot and diffuse and hardly contributes to the BH evolution. The accretion of the cold star-forming gas onto the central BH adds constructively in terms of AM, and over time increases the BH spin to high values even though the turbulence in the ISM is as large as $50 \, \rm km\, s^{-1}$.

When SN feedback activity is at minimum, the mass growth of the BH and its spin evolution are driven by Eddington-limited phases where the gas around the BH is dense and its rotation is imposed by the global rotation of the galactic disc. The phases of random chaotic accretion that could efficiently lower the BH spin are synchronous with periods of SNe explosions that quench the accretion rate onto the BH by creating low-density and hot expanding bubbles.

The cosmological perturbations induced by galaxy mergers and cold filaments can significantly alter the AM of gas in the galaxy and the accretion flow on the BH, reorient the direction of the spin, and lower its value. However, these events are sufficiently rare and short in duration, even at high redshift,  that the BH  spin rapidly returns to its steady-state, maximal value through accretion of this fresh ISM gas. 

The large values of the spin measured in these simulations for gas-rich galaxies have potentially severe consequences on the growth of high-redshift supermassive BH as observed in the SDSS~\citep[e.g.][]{fanetal06}.
Indeed, the characteristic time-scale for the growth of BH is inversely proportional to the radiative efficiency of the accretion process, which, for radiatively efficient accretion discs, is an increasing function of the BH spin. For a radiative efficiency of $\epsilon_{\rm r}=0.1$, a value commonly used in cosmological simulations~\citep{sijackietal07, dimatteoetal08, booth&schaye09, duboisetal12agnmodel} and in agreement with Soltan's argument, the e-folding Salpeter time is $t_{\rm salp}=45 \, \rm Myr$, meaning that BH can grow as large as $10^9\, \rm M_\odot$ within $1\, \rm Gyr$ ($z=6$) even when starting from large stellar seed mass values ($10^2-10^3 \, \rm M_\odot$).
For BH spinning with values $a=0.998$ as in our simulations, the radiative efficiency and the Salpeter time become $\epsilon_{\rm r}=0.32$ and $t_{\rm salp}=135 \, \rm Myr$. Thus, a maximally spinning BH cannot reach $10^9\, \rm M_\odot$ in less than a Gyr if its accretion is Eddington-limited.
One simple way to solve this apparent mismatch is that BH grow at super-Eddington rates \citep[e.g.,][]{volonterirees05}. Super-Eddington accretion alleviates the time-scale constraint provided that there is enough gas in the very centre of galaxies early on, which seems to be the case~\citep{duboisetal12angmom}. At the same time, in radiatively inefficient super-Eddington accretion discs (e.g., slim discs, but note that this is true for the case of very sub-Eddington flows as well) there is no expectation that the radiative efficiency strongly depends on BH spin, at variance with mildly sub-Eddington radiatively efficient discs, such as Shakura-Sunyaev discs.

This study of BH spin evolution in a realistic hydrodynamical cosmological environment is a first step towards future more resolved and more realistic simulations (including radiation from BHs and from stars, for instance), and more systematic studies by varying halo mass and halo histories for which the level of turbulence in the ISM and its coherence on large galactic scales could differ by large amounts (e.g. in discs, ellipticals, clumpy high-redshift galaxies).

\section*{Acknowledgments}
We thank the anonymous referee for suggestions, which improved the clarity of the paper.
The simulations presented here were run on the DiRAC-1 and DiRAC-2 facilities jointly funded by STFC, the Large Facilities Capital Fund of BIS and the Universities
 of Oxford and Leicester. 
This research is part of the Horizon-UK project.
This research has been supported in part by the International Balzan Foundation via New College, Oxford, and by the National Science Foundation under Grant No. NSF PHY11-25915.
YD and JS acknowledge support by the ERC advanced grant (Dark Matters).
MV acknowledges funding support from NASA, through award ATP NNX10AC84G, and from a Marie Curie Career Integration grant (PCIG10-GA-2011-303609). 

\bibliographystyle{mn2e}
\bibliography{author}

\appendix

\section{Resolution study}
\label{appendix:resolution}

In this section, we test how the results depend on the simulations' resolution, by changing the minimum cell size (maximum level of refinement) in the isolated galaxy run. We performed three simulations at lower resolution $\Delta x=20$, $40$ and $80\, \rm pc$. In the three additional runs the density threshold for star formation in the polytropic EoS, $n_0=62.5$, $15.5$ and $4 \, \rm H\, cm^{-3}$, and the dissipation time-scale of the non-thermal component for SNe feedback $t_{\rm diss}=0.95$, $2.41$ and $4.82\, \rm Myr$ scale with the resolution. These parameters are calibrated to keep the propagation length of the blast wave constant in number of resolution elements ($\lambda_{\rm Jeans}=4 \Delta x$) for explosions occurring at $n=n_0$.

\begin{figure*}
  \centering{\resizebox*{!}{4.25cm}{\includegraphics{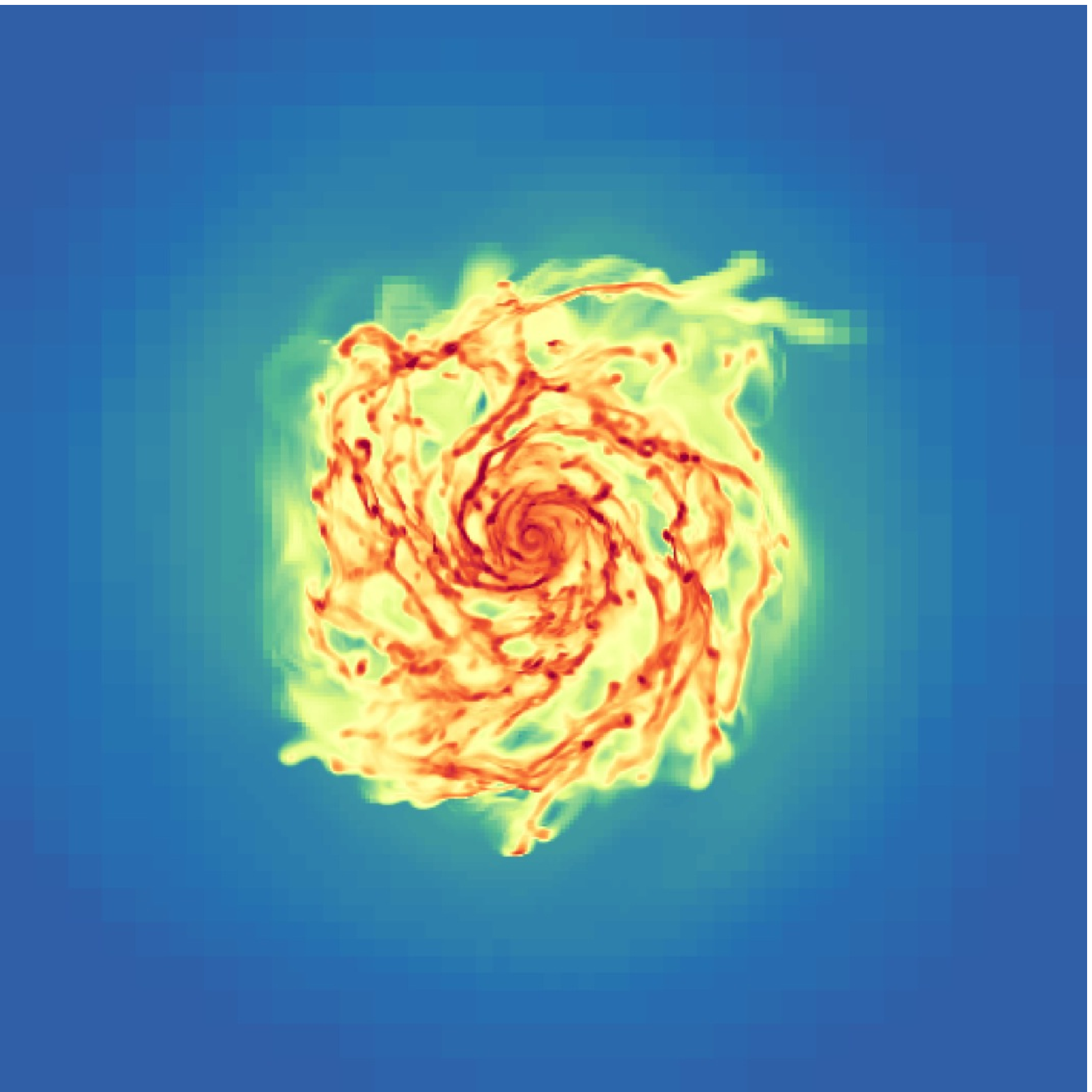}}}
  \centering{\resizebox*{!}{4.25cm}{\includegraphics{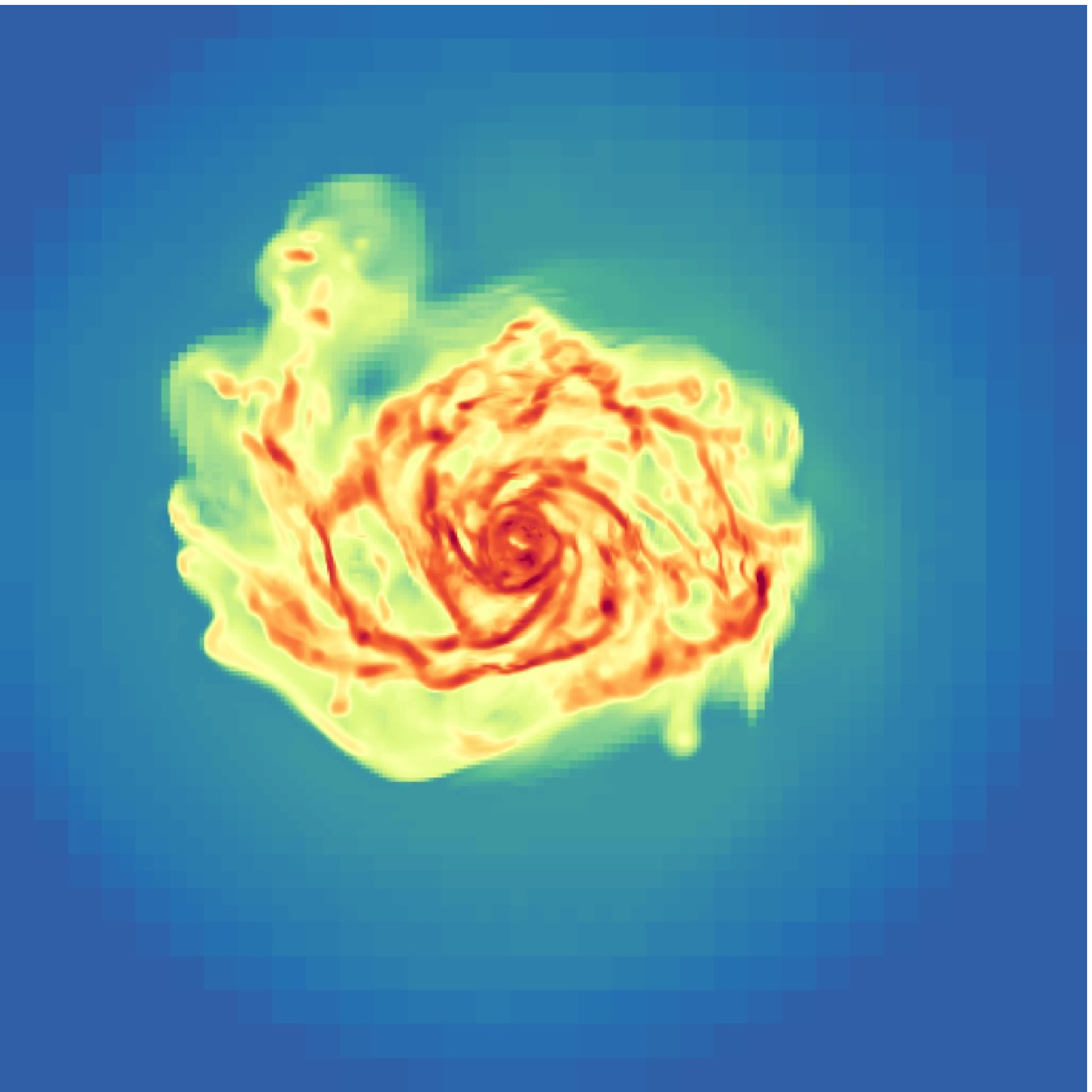}}}
  \centering{\resizebox*{!}{4.25cm}{\includegraphics{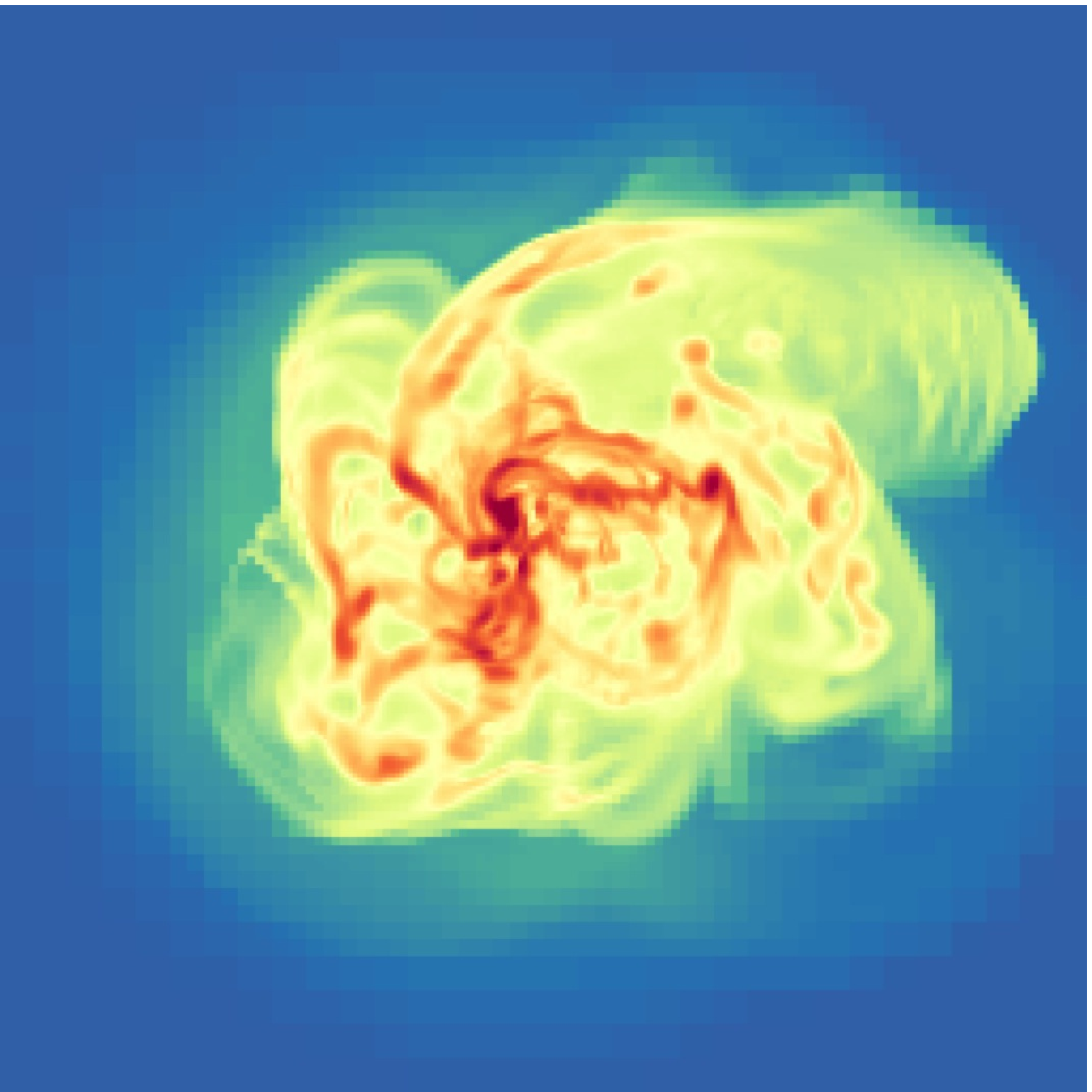}}}
  \centering{\resizebox*{!}{4.25cm}{\includegraphics{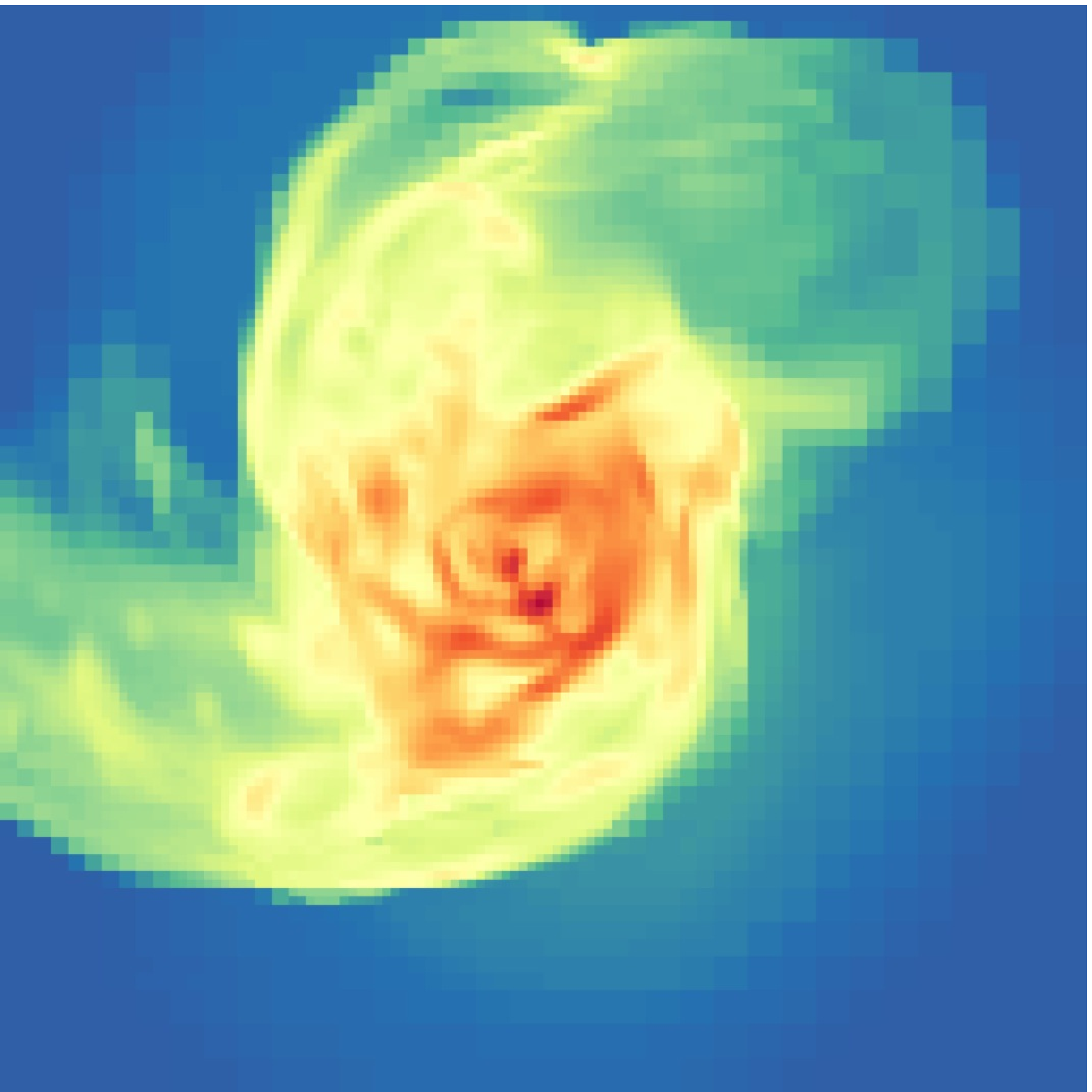}}}
  \centering{\resizebox*{!}{4.25cm}{\includegraphics{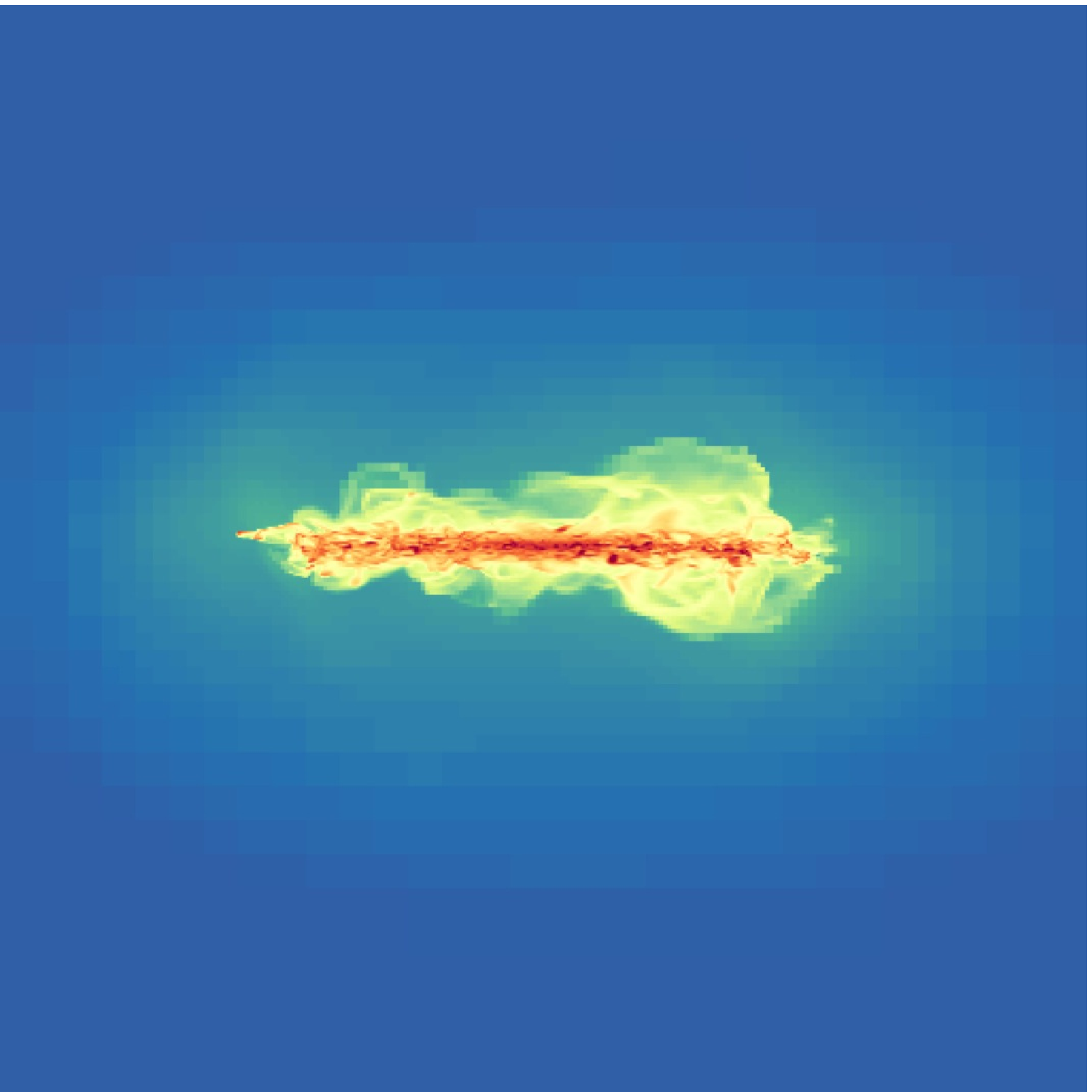}}}
  \centering{\resizebox*{!}{4.25cm}{\includegraphics{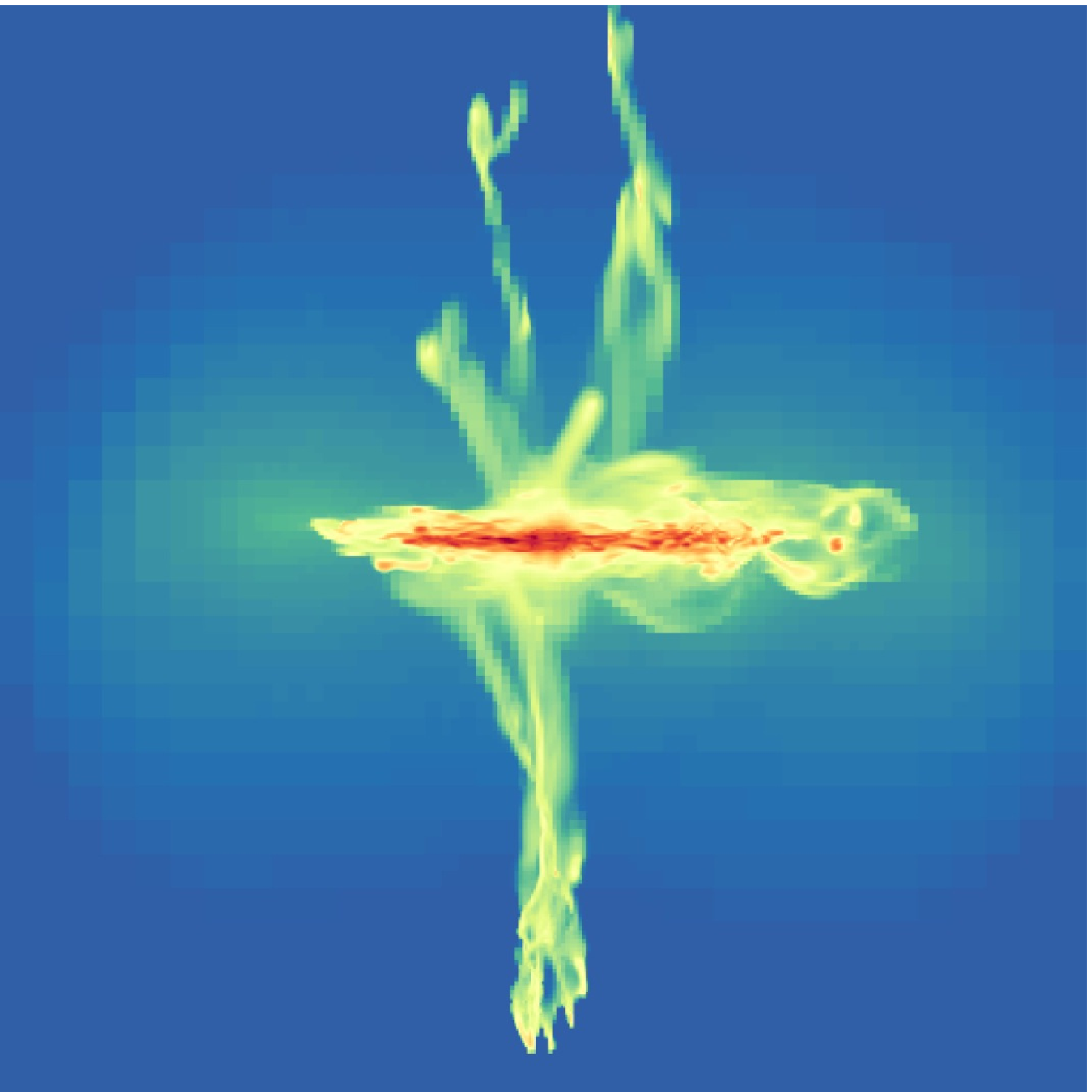}}}
  \centering{\resizebox*{!}{4.25cm}{\includegraphics{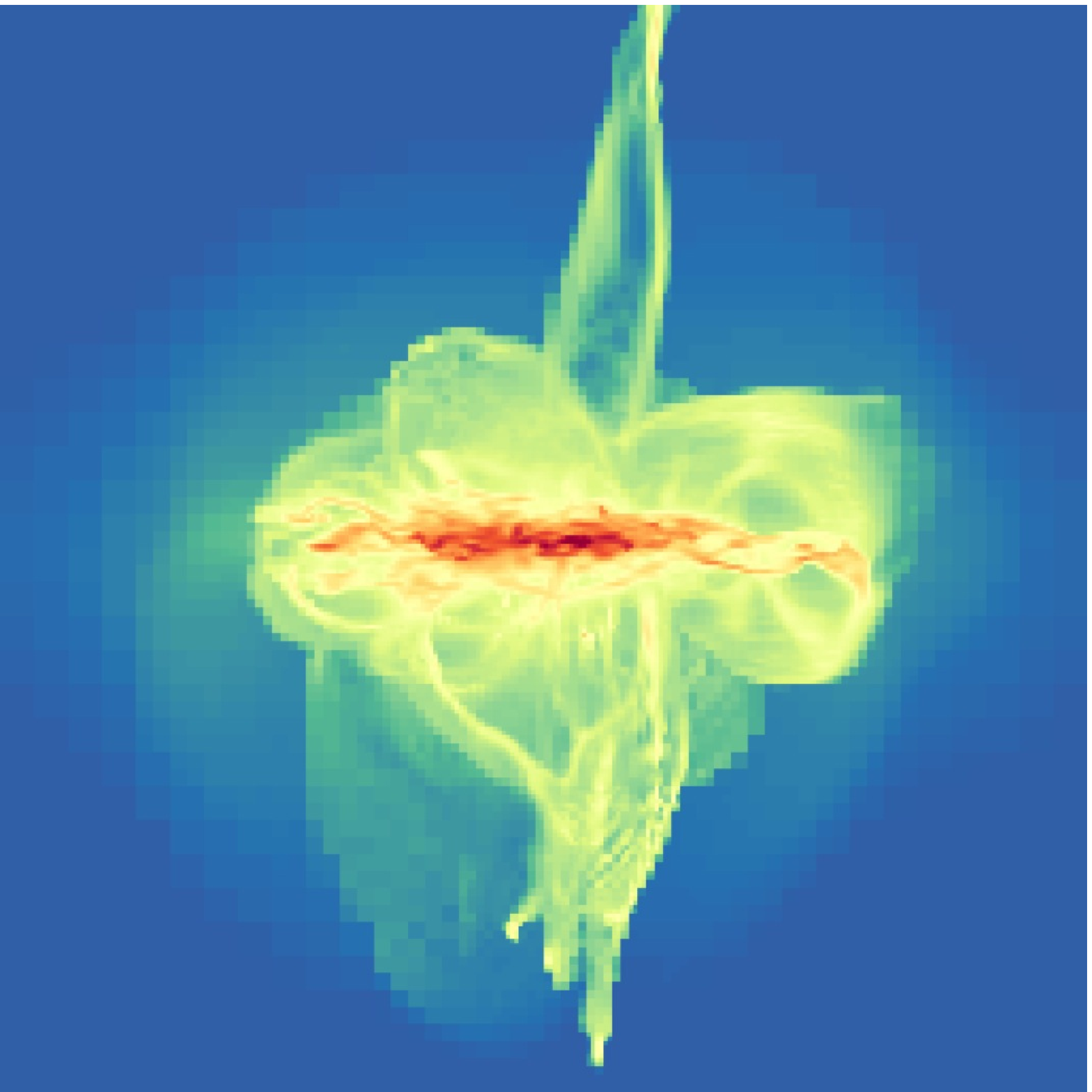}}}
  \centering{\resizebox*{!}{4.25cm}{\includegraphics{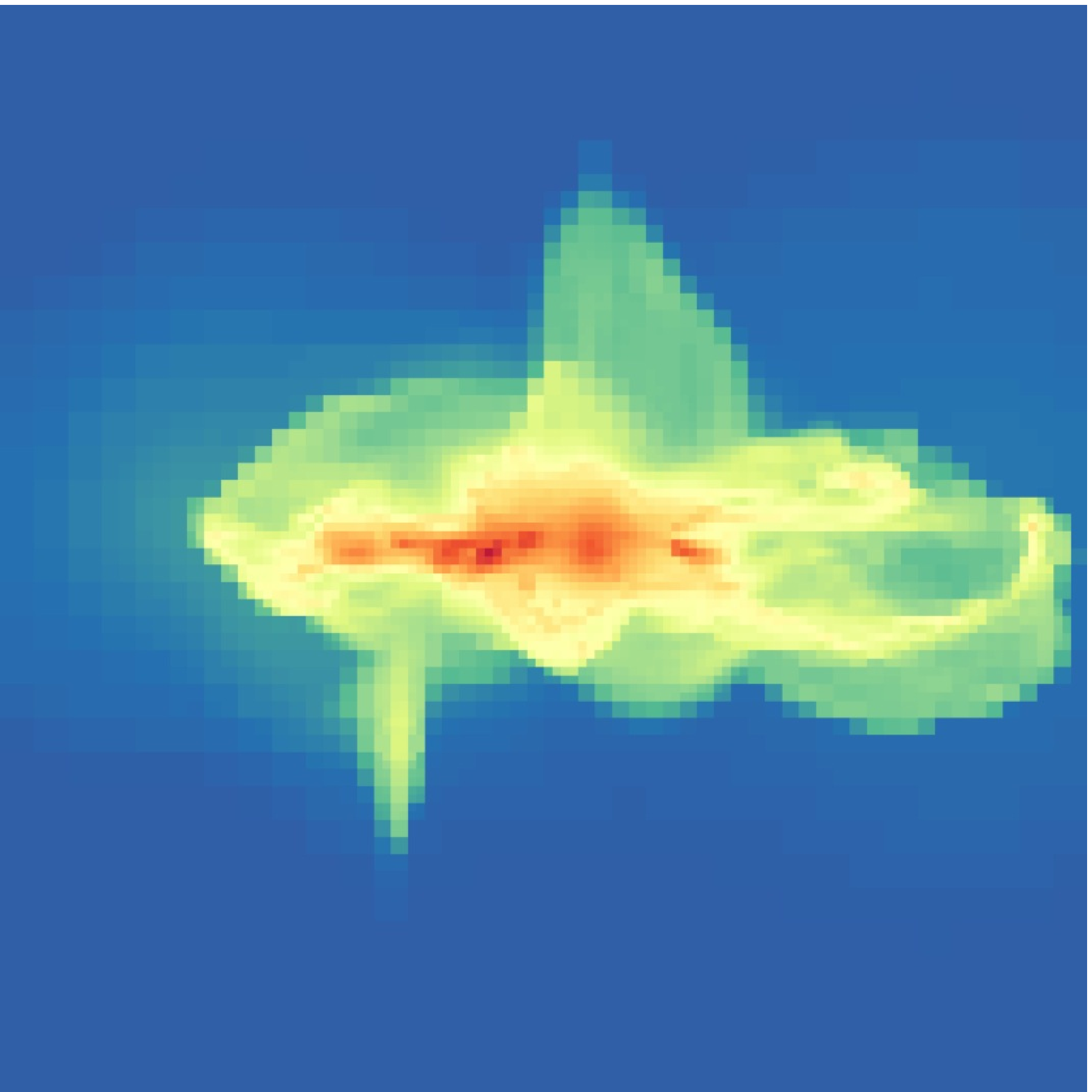}}}
  \caption{Face-on views (top panels) and edge-on views (bottom panels) of the gas density of the galaxy at $t=800$ Myr for different resolutions $\Delta x=10$~pc (left columns), $\Delta x=20$~pc (middle left columns), $\Delta x=40$~pc (middle right columns), $\Delta x=80$~pc (right columns). The same color scaling is applied for all images and images are 10 kpc wide.}
    \label{fig:gdensresol}
\end{figure*}

\begin{figure*}
  \centering{\resizebox*{!}{4.cm}{\includegraphics{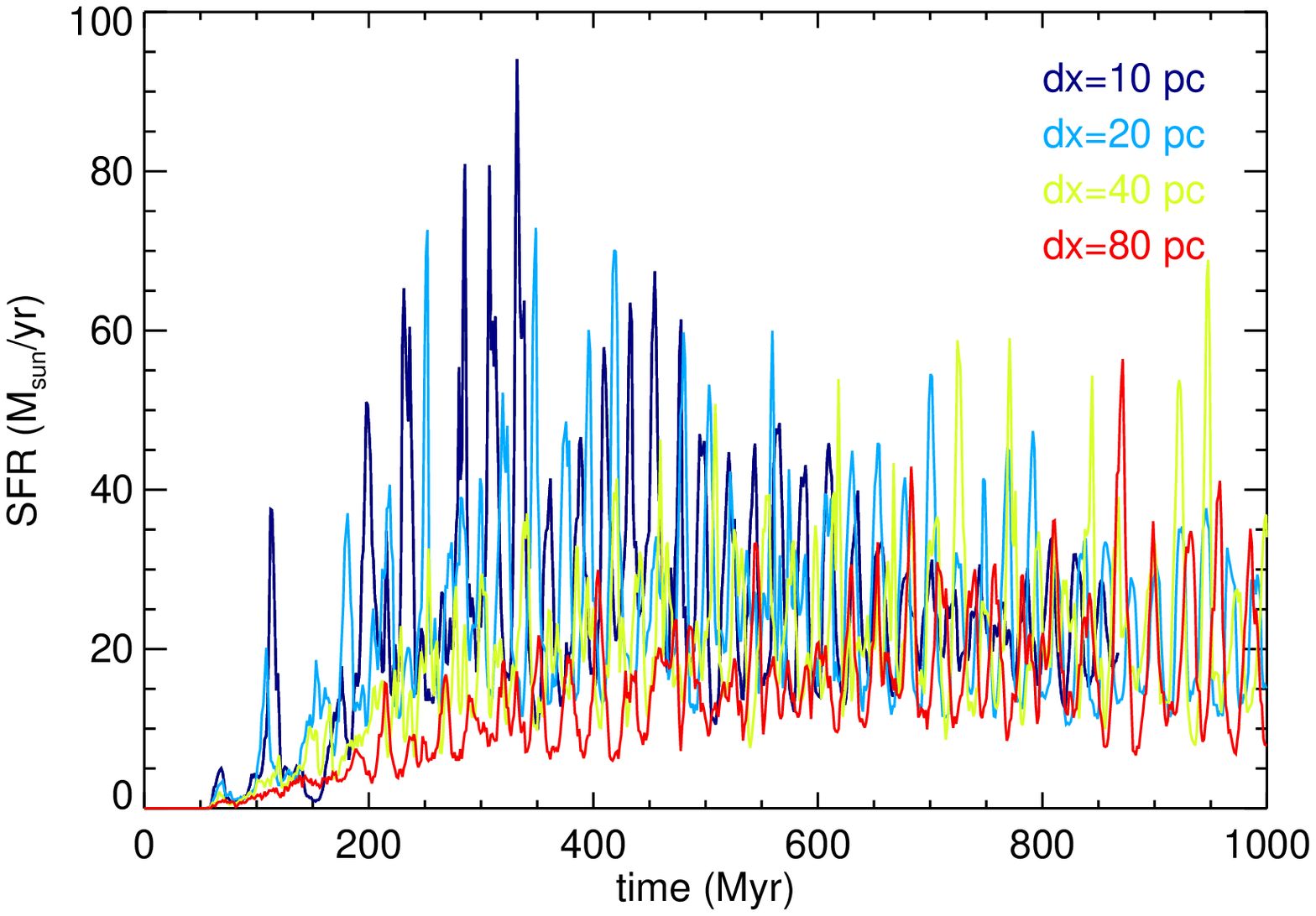}}}
  \centering{\resizebox*{!}{4.cm}{\includegraphics{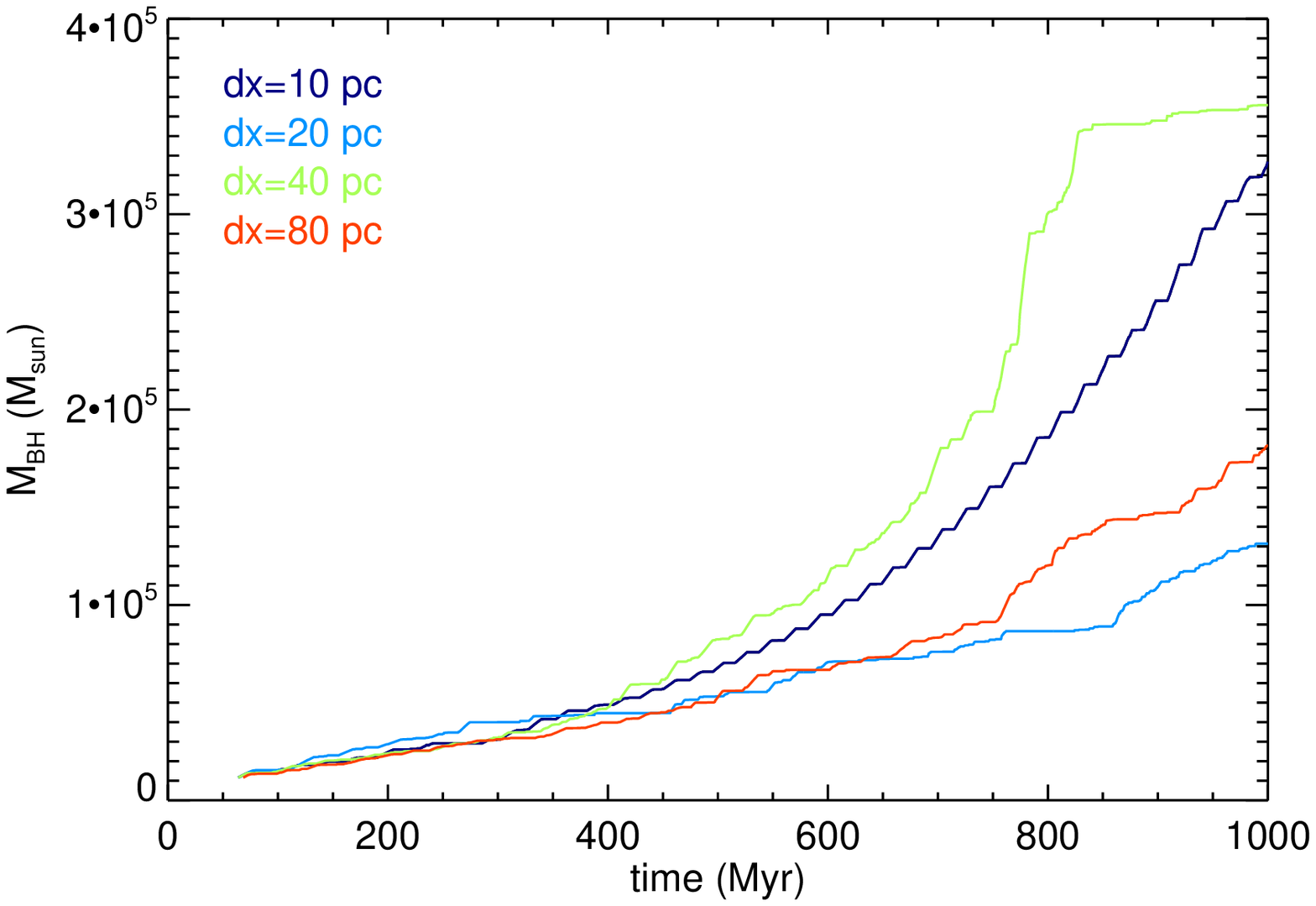}}}
  \centering{\resizebox*{!}{4.cm}{\includegraphics{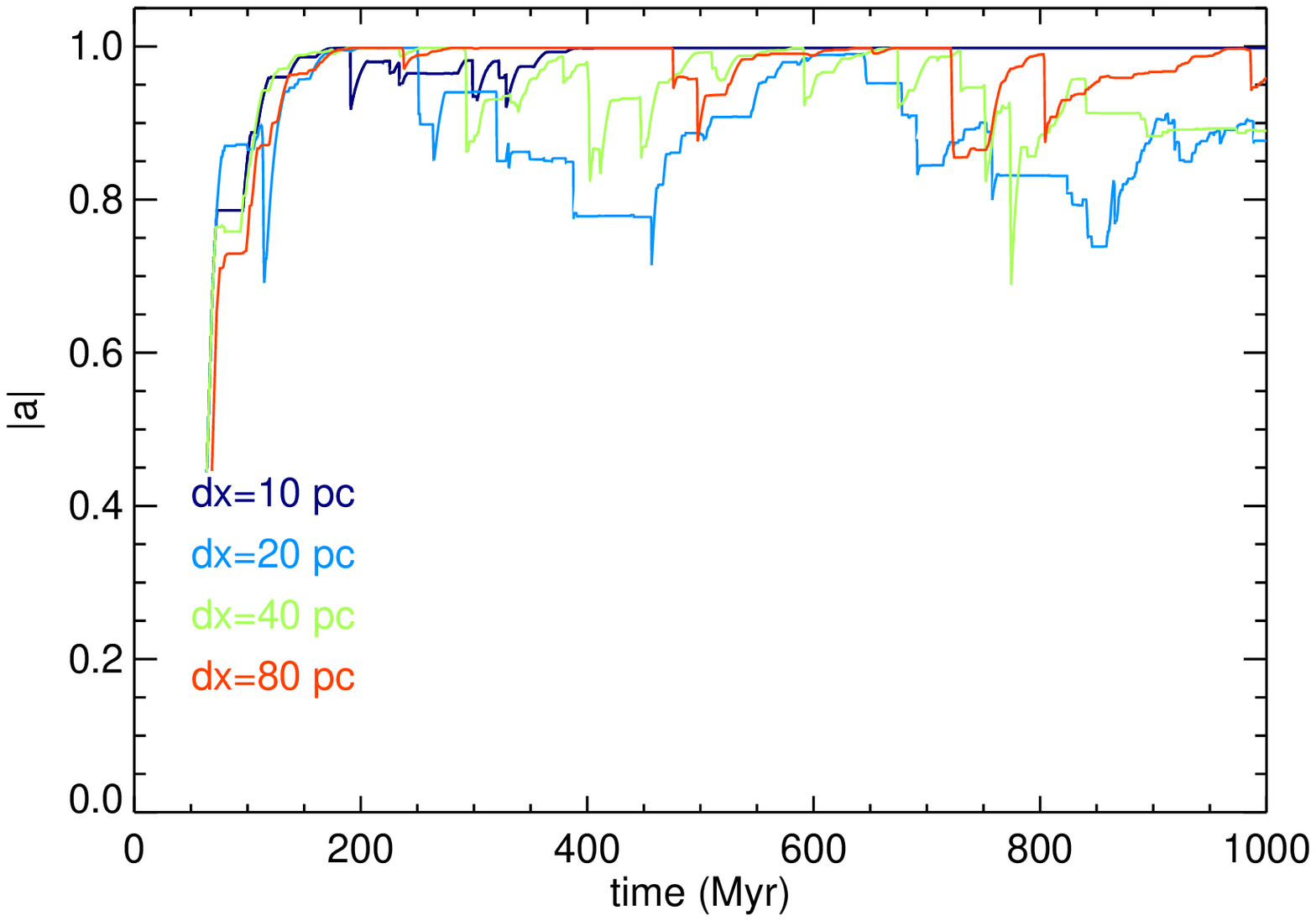}}}
  \caption{Star formation rate (left panel), BH mass (middle panel) and spin (right panel) as a function of time for different spatial resolutions 10, 20, 40 and 80 pc for the isolated halo initial conditions.}
    \label{fig:evolresol}
\end{figure*}

Projections of the face-on and edge-on gas densities are shown in Fig.~\ref{fig:gdensresol} after 800 Myr of evolution.
As expected, the small-scale structure of the ISM is smoothed out with decreasing resolution.
Large-scale features such as gas concentrations in the central region, spiral arms of gas and 100 pc SNe-driven bubble are however  observed in all four runs with different resolutions.
The consequence is that the average star formation rates are similar $<{SFR}>\simeq 20 \, \rm M_\odot \, yr^{-1}$ but the bursts appear at different times and with different amplitudes (left panel of Fig.~\ref{fig:evolresol}).
As the details of star formation change, the fuelling of the central BH differ and the central supermassive BH  has a slightly different mass growth and spin evolution. 
However, the final BH masses end up with similar values between $1$-$3\times 10^6 \, \rm M_\odot$  (middle panel of Fig.~\ref{fig:evolresol}) and similar values of the spin are observed in all runs, with $\vert a\vert\gtrsim0.8$ (right panel of Fig.~\ref{fig:evolresol}). This  resolution study shows that BHs grow at about Eddington with large spin values independently of how 
much the multiphase structure of the ISM is resolved.

\section{A different modeling of Supernova feedback}
\label{appendix:snkinetic}

In this Appendix, we present the result of the isolated galaxy simulation with $\Delta x=10\, \rm pc$ and a different scheme for SN feedback  described in~\cite{dubois&teyssier08winds}.
This feedback modelling is based on a kinetic approach where mass, momentum and energy are released in the immediate vicinity of the exploding stellar particle, and cooling takes place everywhere and at all times, i.e. independently of whether recent explosions have occurred in a specific place or not.
With this approach, SN feedback becomes effective only in diffuse regions of the ISM, as energy is quickly thermalised and radiated away when the explosions take place in cold dense clouds.
As a consequence, star-forming clouds are long-lived structures, until they consume all their gas into stars or are captured by the central bulge.
\cite{tasker11} have shown that amongst the large variety of star-forming clouds in Toomre unstable discs, there are co- and counter-rotating clumps (with a preference for co-rotating clumps). 

Therefore, with that feedback implementation, the cold gas feeds the central BH in a coherent fashion up to the point where this BH gets massive enough to self-regulate and blow away the gas in its immediate vicinity. Once the self-regulated state is achieved, only the cold star-forming clouds that manage to reach the very centre of the galaxy trigger episodes of massive fuelling of the BH ~\citep{bournaudetal11, duboisetal12angmom, duboisetal13, duboisetal13spinlss, gabor&bournaud13, bournaudetal13}. 

\section{Spin alignment timescales}
\label{appendix:spinalign}

The timescale over which BH spins align with the angular momentum of the accretion flow, and its relation to the accretion timescale have been the subject of a long debate for decades \citep[e.g.,][]{1975ApJ...195L..65B,1978Natur.275..516R,1983MNRAS.202.1181P,1996MNRAS.282..291S,NatarajanPringle1998,2000MNRAS.315..570N,2005ApJ...623..347F,2007ApJ...668..417F,2011ApJ...730...36D,2013ApJ...768..133S,sorathia2013}.  In this Appendix we explore the consequences of alignment timescales longer than assumed in the main article, and we test how sensitive the results are to the accretion rate behaviour and its synchronisation with gas cooling. We analyse the isolated galaxy run, and we post-process the relative BH alignment and spin magnitude, considering as input the direction of the accreted gas angular momentum, and the accretion rate. This analysis is therefore simplified with respect to the main text, where spin is tracked on-the-fly, but allows us to highlight the main dependences in a clear way. 
\begin{figure*}
  \centering{\resizebox*{!}{\columnwidth}{\includegraphics{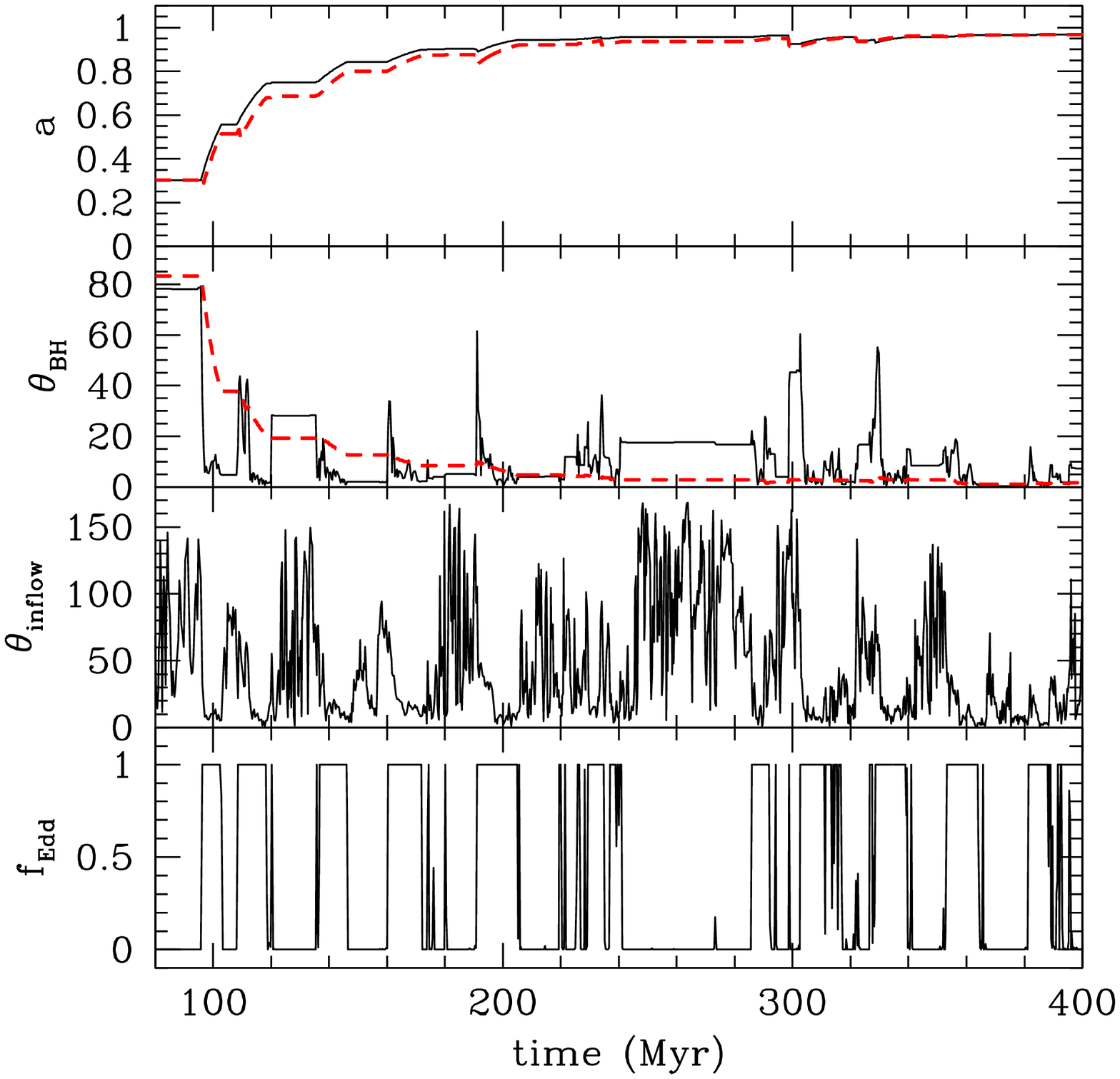}}}
  \centering{\resizebox*{!}{\columnwidth}{\includegraphics{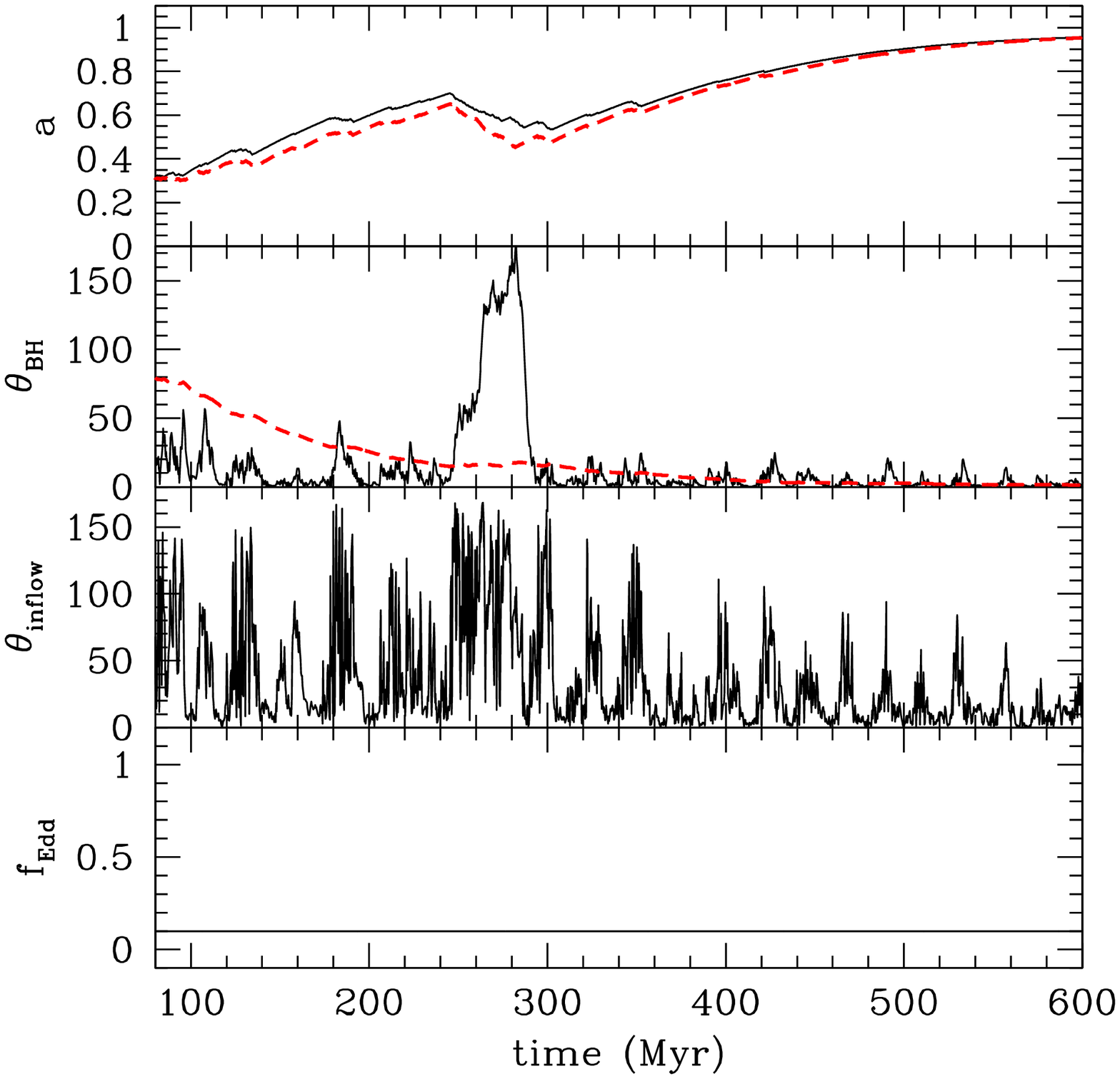}}}
  \caption{Evolution of BH properties in the isolated galaxy run as a function of time. \emph{Left panel}:  we post-process the relative BH alignment and spin magnitude, considering as input the direction of the accreted gas angular momentum and the accretion rate. \emph{Right panel}: the accretion rate is set to 10 per cent of the Eddington rate. The black solid curves track spin magnitude and direction assuming fast spin alignment; the red dashed curves assume instead slow spin alignment.}
    \label{fig:spinalign}
\end{figure*}

In Figure~\ref{fig:spinalign} (left) we show the time evolution of the BH spin, angle of the BH spin vector and of the inflow angular momentum with respect to the z-axis,  and accretion rate for the first 400 Myr of the galaxy evolution (the ``galactic fountain'' phase, cf. Fig.~\ref{fig:outflow}).  As discussed in the main text, high-accretion rates are synchronised with gas cooling and formation of a central rotationally-supported structure, and this is apparent when comparing the two bottom panels. The top two panels show spin magnitude and direction. The black solid curves assume fast alignment, while the red dashed curves show a case with slower alignment  ($\nu_2=\nu_1$). One can notice the different behaviour of the spin direction with time, but the spin magnitude is hardly modified. This is because in the phases of maximum chaos the accretion rate on the BH is low, and therefore the spin magnitude is not affected. In Figure~\ref{fig:spinalign} (right) we perform an additional test, where we artificially set the accretion rate to 10 per cent of the Eddington rate at all times, while keeping the information on the direction of the inflow (the third panel from top is the same in both plots). In this experiment, we can assess the effect of prolonged phases of chaotic accretion at high rates (which do not occur in our simulations, due to the synchronisation between cooling, disc formation and BH feeding).  Here, in the case of fast alignment (black solid curves) we see that the most chaotic phases of accretion give rise to occasional counter-rotating accretion (e.g., between 250 and 300 Myr) that decrease the spin magnitude. We extend this plot to 600 Myr, where the galaxy reaches a more quiescent regime (``hot bubbles" phase), and a ``secular" evolution of the angular momentum direction is reached. By this time, regardless of the assumptions on alignment timescales, the BH spin reaches a steady state at high values. 

In conclusion, the alignment timescale is secondary to the dynamics of accretion in determining the BH spin magnitude evolution. However, it is the main driver of how fast the spin direction changes, regardless of the accretion rate. 

\end{document}